\PassOptionsToPackage{unicode}{hyperref}
\PassOptionsToPackage{hyphens}{url}
\PassOptionsToPackage{dvipsnames,svgnames,x11names}{xcolor}
\documentclass[
  12pt,
  letterpaper,
]{article}
\usepackage{xcolor}
\usepackage[margin=1in]{geometry}
\usepackage{amsmath,amssymb}
\usepackage{iftex}
\ifPDFTeX
  \usepackage[T1]{fontenc}
  \usepackage[utf8]{inputenc}
  \usepackage{textcomp} % provide euro and other symbols
\else % if luatex or xetex
  \usepackage{unicode-math} % this also loads fontspec
  \defaultfontfeatures{Scale=MatchLowercase}
  \defaultfontfeatures[\rmfamily]{Ligatures=TeX,Scale=1}
\fi
\usepackage{lmodern}
\ifPDFTeX\else
\fi
\IfFileExists{upquote.sty}{\usepackage{upquote}}{}
\IfFileExists{microtype.sty}{% use microtype if available
  \usepackage[]{microtype}
  \UseMicrotypeSet[protrusion]{basicmath} % disable protrusion for tt fonts
}{}
\makeatletter
\@ifundefined{KOMAClassName}{% if non-KOMA class
  \IfFileExists{parskip.sty}{%
    \usepackage{parskip}
  }{% else
    \setlength{\parindent}{0pt}
    \setlength{\parskip}{6pt plus 2pt minus 1pt}}
}{% if KOMA class
  \KOMAoptions{parskip=half}}
\makeatother
\makeatletter
\ifx\paragraph\undefined\else
  \let\oldparagraph\paragraph
  \renewcommand{\paragraph}{
    \@ifstar
      \xxxParagraphStar
      \xxxParagraphNoStar
  }
  \newcommand{\xxxParagraphStar}[1]{\oldparagraph*{#1}\mbox{}}
  \newcommand{\xxxParagraphNoStar}[1]{\oldparagraph{#1}\mbox{}}
\fi
\ifx\subparagraph\undefined\else
  \let\oldsubparagraph\subparagraph
  \renewcommand{\subparagraph}{
    \@ifstar
      \xxxSubParagraphStar
      \xxxSubParagraphNoStar
  }
  \newcommand{\xxxSubParagraphStar}[1]{\oldsubparagraph*{#1}\mbox{}}
  \newcommand{\xxxSubParagraphNoStar}[1]{\oldsubparagraph{#1}\mbox{}}
\fi
\makeatother

\usepackage{longtable,booktabs,array}
\usepackage{calc} % for calculating minipage widths
\usepackage{etoolbox}
\makeatletter
\patchcmd\longtable{\par}{\if@noskipsec\mbox{}\fi\par}{}{}
\makeatother
\IfFileExists{footnotehyper.sty}{\usepackage{footnotehyper}}{\usepackage{footnote}}
\makesavenoteenv{longtable}
\usepackage{graphicx}
\makeatletter
\newsavebox\pandoc@box
\newcommand*\pandocbounded[1]{% scales image to fit in text height/width
  \sbox\pandoc@box{#1}%
  \Gscale@div\@tempa{\textheight}{\dimexpr\ht\pandoc@box+\dp\pandoc@box\relax}%
  \Gscale@div\@tempb{\linewidth}{\wd\pandoc@box}%
  \ifdim\@tempb\p@<\@tempa\p@\let\@tempa\@tempb\fi% select the smaller of both
  \ifdim\@tempa\p@<\p@\scalebox{\@tempa}{\usebox\pandoc@box}%
  \else\usebox{\pandoc@box}%
  \fi%
}
\def\fps@figure{htbp}
\makeatother

\NewDocumentCommand\citeproctext{}{}
\NewDocumentCommand\citeproc{mm}{%
  \begingroup\def\citeproctext{#2}\cite{#1}\endgroup}
\makeatletter
 \let\@cite@ofmt\@firstofone
 \def\@biblabel#1{}
 \def\@cite#1#2{{#1\if@tempswa , #2\fi}}
\makeatother
\newlength{\cslhangindent}
\newlength{\csllabelwidth}
\newenvironment{CSLReferences}[2] % #1 hanging-indent, #2 entry-spacing
 {\begin{list}{}{%
  \setlength{\itemindent}{0pt}
  \setlength{\leftmargin}{0pt}
  \setlength{\parsep}{0pt}
  \ifodd #1
   \setlength{\leftmargin}{\cslhangindent}
   \setlength{\itemindent}{-1\cslhangindent}
  \fi
  \setlength{\itemsep}{#2\baselineskip}}}
 {\end{list}}
\usepackage{calc}

\providecommand{\tightlist}{%
  \setlength{\itemsep}{0pt}\setlength{\parskip}{0pt}}

\usepackage{setspace}
\usepackage{booktabs}
\usepackage{etoolbox}
\AtBeginEnvironment{longtable}{\singlespacing}
\AtBeginEnvironment{tabular}{\singlespacing}
\usepackage{threeparttable}
\usepackage{caption}
\usepackage{multirow}
\usepackage{pdflscape}
\usepackage{tikz}
\usetikzlibrary{arrows.meta, positioning, calc}

\makeatletter
\@ifpackageloaded{caption}{}{\usepackage{caption}}
\AtBeginDocument{%
\ifdefined\contentsname
  \renewcommand*\contentsname{Table of contents}
\else
  \newcommand\contentsname{Table of contents}
\fi
\ifdefined\listfigurename
  \renewcommand*\listfigurename{List of Figures}
\else
  \newcommand\listfigurename{List of Figures}
\fi
\ifdefined\listtablename
  \renewcommand*\listtablename{List of Tables}
\else
  \newcommand\listtablename{List of Tables}
\fi
\ifdefined\figurename
  \renewcommand*\figurename{Figure}
\else
  \newcommand\figurename{Figure}
\fi
\ifdefined\tablename
  \renewcommand*\tablename{Table}
\else
  \newcommand\tablename{Table}
\fi
}
\@ifpackageloaded{float}{}{\usepackage{float}}
\floatstyle{ruled}
\@ifundefined{c@chapter}{\newfloat{codelisting}{h}{lop}}{\newfloat{codelisting}{h}{lop}[chapter]}
\floatname{codelisting}{Listing}

\makeatother
\makeatletter
\@ifpackageloaded{caption}{}{\usepackage{caption}}
\@ifpackageloaded{subcaption}{}{\usepackage{subcaption}}
\makeatother
\usepackage{bookmark}
\IfFileExists{xurl.sty}{\usepackage{xurl}}{} % add URL line breaks if available
\makeatletter
\@ifundefined{xmpquote}{}{}
\makeatother
\hypersetup{
  pdftitle={Non-Proportionality{,} Structural Persistence{,} and the Bifactor Decision},
  pdfauthor={Jinsong Chen Faculty of Education, The University of Hong Kong (jinsong.chen@live.com)},
  colorlinks=true,
  linkcolor={blue},
  filecolor={Maroon},
  citecolor={Blue},
  urlcolor={Blue},
  pdfcreator={LaTeX via pandoc}}

\title{Non-Proportionality, Structural Persistence, and the Bifactor
Decision}
\author{Jinsong Chen \tabularnewline Faculty of Education, The
University of Hong Kong (jinsong.chen@live.com)}
\date{}
\begin{document}
\maketitle
\begin{abstract}
Whether an added general dimension is necessary beyond correlated
first-order factors is a property of population covariance, not an
estimator. A bifactor structure is covariance-equivalent to correlated
factors when general and group loadings are proportional within every
cluster. When every cluster violates proportionality, at least three
indicators per cluster and mild conditions rule out exact reproduction
by a \(K\)-factor model with diagonal uniquenesses. Mixed configurations
remain only partly characterized, motivating graded distinguishability.
We develop a two-step procedure that delivers a stable first-order
structure only when it persists across direct wider-count comparisons,
then compares oblique and bifactor representations conditionally on
delivery. A preliminary simulation study provides tentative design
evidence. A major study develops three non-nested persistence profiles,
designates \texttt{.80/r2} as the practical default, and evaluates the
frozen family on held-out replications from the same generators. Within
the studied scenarios, anchor-only persistence outperforms anchor-zero,
minimum congruence modestly outperforms maximum RMSD, and depth four
adds no precision at the fixed .70 cutoff. Held-out results also expose
stable-yet-wrong delivery under a weak core and show that the
conditional comparison is highly accurate after correct recovery. Four
empirical applications illustrate agreement, adjacent-count uncertainty,
and non-delivery. Persistence--not count alone--carries the structural
decision.

\emph{Keywords:} bifactor model; distinguishability; covariance
equivalence; partially exploratory factor analysis; structural
persistence
\end{abstract}

\section{Introduction}\label{sec-intro}

Few measurement models are as attractive and as distrusted as the
bifactor model. Since its rediscovery
(\citeproc{ref-reise2012rediscovery}{Reise, 2012}), it has been fit to
ability batteries, psychopathology inventories, and quality-of-life
scales. It also wins the model comparisons it enters with remarkable
regularity: more than 90\% of the time against the higher-order model in
mental-abilities batteries (\citeproc{ref-cucina2017bifactor}{Cucina \&
Byle, 2017}). Yet the same literature documents chronic anomalies, with
a reported prevalence above 60\% in applications
(\citeproc{ref-eid2017anomalous}{Eid et al., 2017}). These include
collapsed or vanishing group factors, general factors that rotate into
group factors, and loading patterns that defy interpretation.
Methodologists have raised three further concerns. The bifactor's fit
advantage partly reflects its greater functional flexibility rather than
structural truth (\citeproc{ref-bonifay2017complexity}{Bonifay \& Cai,
2017}; \citeproc{ref-murray2013limitations}{Murray \& Johnson, 2013}).
The higher-order model is the bifactor plus a proportionality constraint
(\citeproc{ref-gignac2016higher}{Gignac, 2016};
\citeproc{ref-schmid1957development}{Schmid \& Leiman, 1957};
\citeproc{ref-yung1999relationship}{Yung et al., 1999}). Standard
information criteria can prefer the wrong member of the pair at
realistic sample sizes (\citeproc{ref-raykov2024bic}{Raykov et al.,
2024}).

This paper starts from one observation. These troubles are not
independent pathologies. They are manifestations of one covariance-level
fact: \emph{whether a general factor is empirically distinguishable from
correlated factors is a property of the population loading pattern}.
Specifically, it depends on the proportionality between general and
group loadings within clusters. When the pattern is proportional in
every cluster, the bifactor covariance is exactly reproducible by a
higher-order model, and the two classes are covariance-equivalent even
when the parameters remain identified within a chosen class. When every
cluster departs from proportionality, the classes are distinguishable at
the covariance level (Theorem 1); a departure confined to part of the
battery is not enough in general. Most real data sit somewhere between
these poles, and \emph{where} they sit determines how much an analyst
can safely leave to the data.

Two literatures approach this problem from opposite identification
philosophies, and both descend from Schmid and Leiman
(\citeproc{ref-schmid1957development}{1957}). The exploratory lineage
identifies an unrestricted loading matrix by rotation or constraint
(\citeproc{ref-asparouhov2026unification}{Asparouhov \& Muthén, 2026};
\citeproc{ref-garciagarzon2019improving}{Garcia-Garzon et al., 2019};
\citeproc{ref-jennrich2011exploratory}{Jennrich \& Bentler, 2011};
\citeproc{ref-jimenez2023exploratory}{Jiménez et al., 2023};
\citeproc{ref-qiao2025exact}{Qiao et al., 2025a}): the analyst supplies
only the factor count, anchor knowledge never enters estimation, and
interpretation is post hoc. The partially exploratory lineage starts
from the confirmatory pole: a few anchors are estimated structure rather
than rotation targets, the remaining pattern is selected by
regularization, and candidate structures grow from an anchored backbone
through a factor-structure sweep (\citeproc{ref-chenjin2026fit}{Chen \&
Jin, 2026a}). The positioning difference is what each delivers.
Exploratory hierarchical approaches condition on a supplied count and
ask which representation is recoverable at that count. The present
approach treats count uncertainty as part of the problem: its target is
the stable structural core that persists across direct comparisons to
larger candidates in a defensible window, with no stable structure
delivered as the explicit output when none exists, and the bifactor
decision conditional on the delivered structure
(Table~\ref{tbl-positioning}). Relative to the framework's own
factor-selection machinery (\citeproc{ref-chenjin2026fit}{Chen \& Jin,
2026a}), the increments are the distinguishability theory, the
persistence of stable structure as the delivery estimand, with declared
depth, explicit non-delivery, and a frozen family of persistence
profiles in place of a single count rule, the held-out end-to-end
evaluation, and the recast of the bifactor decision as a necessity
question.

\begin{longtable}[]{@{}
  >{\raggedright\arraybackslash}p{(\linewidth - 4\tabcolsep) * \real{0.3134}}
  >{\raggedright\arraybackslash}p{(\linewidth - 4\tabcolsep) * \real{0.3433}}
  >{\raggedright\arraybackslash}p{(\linewidth - 4\tabcolsep) * \real{0.3433}}@{}}
\caption{The positioning contrast: what is conditioned on and what is
delivered.}\label{tbl-positioning}\tabularnewline
\toprule\noalign{}
\begin{minipage}[b]{\linewidth}\raggedright
Approach
\end{minipage} & \begin{minipage}[b]{\linewidth}\raggedright
Treatment of the factor count
\end{minipage} & \begin{minipage}[b]{\linewidth}\raggedright
Primary deliverable
\end{minipage} \\
\midrule\noalign{}
\endfirsthead
\toprule\noalign{}
\begin{minipage}[b]{\linewidth}\raggedright
Approach
\end{minipage} & \begin{minipage}[b]{\linewidth}\raggedright
Treatment of the factor count
\end{minipage} & \begin{minipage}[b]{\linewidth}\raggedright
Primary deliverable
\end{minipage} \\
\midrule\noalign{}
\endhead
\bottomrule\noalign{}
\endlastfoot
Rotation- or constraint-based hierarchical EFA
(\citeproc{ref-asparouhov2026unification}{Asparouhov \& Muthén, 2026};
\citeproc{ref-jennrich2011exploratory}{Jennrich \& Bentler, 2011};
\citeproc{ref-qiao2025exact}{Qiao et al., 2025a}) & supplied, or
selected outside the structural analysis & a hierarchical representation
conditional on \(K\) \\
This paper, step 1 & uncertain, examined over a defensible window & a
stable structural core under direct wider-count comparisons, or no
stable structure delivered \\
This paper, step 2 & conditional on the delivered structure & evidence
on whether an additional common dimension is necessary \\
\end{longtable}

Separating the count from the structure matters in both directions. An
accurate count reading is not yet a deliverable structure: in the
held-out confirmation the prospectively frozen ELBO path selects the
generating count in 97.4\% of datasets, yet count and persistence still
diverge under interference or a weak core, and the bfi shows the same
dissociation empirically. Conversely, count uncertainty need not mean
structural uncertainty: neighboring candidates can share a stable core,
as the PID-5 data illustrate, and Step 2 then proceeds once at each
distinct count delivered by the declared profiles.

That determination is the paper's subject. In partially confirmatory
factor analysis (PCFA; \citeproc{ref-chen2021scale}{Chen et al., 2021};
\citeproc{ref-chen2021partially}{Chen, 2021};
\citeproc{ref-jin2025regularized}{Jin \& Chen, 2025a}), and in its
partially exploratory extension (PEFA;
\citeproc{ref-chen2023fully}{Chen, 2023};
\citeproc{ref-chenjin2026fit}{Chen \& Jin, 2026a}), the design matrix
\(Q\) carries what the analyst is willing to fix and leaves the rest to
a spike-and-slab prior. PCFA supplies a full \(Q\) over a stated count;
PEFA launches from a \emph{backbone} \(Q\) and sweeps it over a window
of candidate counts. Within either, \emph{specification depth} is the
uniform number of anchor items per factor the backbone carries; this
paper works at depth 2 throughout, and deeper backbones are possible.
Our central claim is that \textbf{the cost of a free loading is
contingent on reducibility}: free loadings are essentially free when the
population is irreducible with adequate loadings, and hazardous exactly
where the population approaches reducibility, has silent clusters
(clusters whose general and group loadings are exactly proportional), or
is probed at the wrong first-order rank or with the wrong structural
frame altogether. The reducible tier and its silent clusters fall
outside the region where fully exploratory recovery is guaranteed
(\citeproc{ref-qiao2025hierarchical}{Qiao et al., 2025b},
\citeproc{ref-qiao2025exact}{2025a}), and no existing study quantifies
what partial specification buys in that excluded region; the Discussion
returns to the neighboring exploratory and identifiability strands.

The paper makes three contributions. First, Section \ref{sec-theory}
gives the covariance-level core: Proposition 1 states that a reducible
bifactor is covariance-equivalent to correlated factors, and Theorem 1
gives a sufficient condition for the converse separation, with the mixed
boundary and its consequences developed there; on this base the two-step
process separates count screening from the operational bifactor
comparison. Second, Section \ref{sec-sims} reports two simulation
studies: a Preliminary Simulation Study with Stages A and B supplies
tentative component findings, and a Major Simulation Study develops the
operating family against generating truth and estimates the
prospectively frozen rules once in a held-out confirmatory stage. Third,
Section \ref{sec-empirical} applies the process to four empirical
datasets spanning the evidence spectrum. The simulation domain is
deliberately bounded, with fixed five-cluster designs, correct anchors,
and designed interference chosen to isolate the reducibility tiers;
Section \ref{sec-discussion} states the corresponding scope limits.

\section{Analytical framework}\label{sec-theory}

\subsection{Models and notation}\label{sec-notation}

Let \(J\) items fall into \(K\) non-overlapping clusters, with cluster
\(k\) containing item set \(I_k\), \(|I_k| = J_k \ge 3\), and
\(K \ge 3\) unless noted. All models are for continuous responses with
uncorrelated residuals;
\(\mathbf{y} = \Lambda \boldsymbol{\eta} + \boldsymbol{\varepsilon}\),
and the common covariance is \(C = \Lambda \Phi \Lambda'\).

Three structures compete. The \textbf{correlated-factors (oblique)
model} has \(\Lambda = A\) of rank \(K\), with each item loading on its
cluster factor, cross-loadings permitted, and factor correlations
\(\Phi\). The \textbf{orthogonal bifactor model} has
\(\Lambda^* = [\mathbf{b}_g, B_s] \in \mathbb{R}^{J \times K^*}\) with
\(K^* = K + 1\) total factors: a general column \(\mathbf{b}_g\) on
which every item loads, and \(K\) group columns with disjoint supports.
All factors are orthonormal, so
\(C = \mathbf{b}_g\mathbf{b}_g' + \sum_k \mathbf{b}_s^k \mathbf{b}_s^{k\prime}\).
The \textbf{higher-order (HO) model} adds a second-order factor with
loadings \(\gamma_k\) over first-order factors. By the Schmid--Leiman
(SL) transformation, the HO model is exactly the orthogonal bifactor
whose loadings satisfy \(b_{g,j} = \lambda_j \gamma_k\) and
\(b_{s,j} = \lambda_j \sqrt{1-\gamma_k^2}\) for every item \(j\) in
cluster \(k\). This is a \textbf{constant within-cluster ratio}
\(b_{s,j}/b_{g,j} = \sqrt{1-\gamma_k^2}/\gamma_k\)
(\citeproc{ref-schmid1957development}{Schmid \& Leiman, 1957};
\citeproc{ref-yung1999relationship}{Yung et al., 1999}).

Specification is encoded in a design matrix following the PCFA
specification (\citeproc{ref-chen2021partially}{Chen, 2021}). Entry
\(q_{jk} = 1\) declares a specified (estimated, anchoring) loading,
\(q_{jk} = 0\) a loading fixed to zero, and \(q_{jk} = -1\) an
unspecified loading assigned a spike-and-slab prior. Two glosses fix
this paper's usage. A loading coded \(-1\) is \emph{free} only in this
specific sense: unspecified in advance and left to regularized
selection, not an unregularized parameter, and an anchor loading is
estimated, not fixed to a numerical value. The \emph{cost} of such
freedom is likewise specific. Preliminary Stage A studies estimation
error at a supplied structure; Preliminary Stage B studies selection,
count error, and its consequences; the major study evaluates the
complete decision route. We write \(Q_0\) for a \textbf{backbone} design
of \(K_0 \le K\) columns used to launch a factor-structure sweep, and
\(Q\) for a full design of \(K\) columns. In bifactor designs the
general column is added internally and fully specified. \(K\) and
\(K_0\) always count \emph{group} (or first-order) factors,
\(K^* = K+1\) counts all factors of the bifactor, and ``the count''
always refers to \(K\); the stable-core count that step 1 targets, the
common-factor rank of \(\Sigma\), and any theoretical cluster count are
distinct quantities and are named separately where they differ.

Three designs recur throughout, ordered by the pattern information they
encode. All three are partial designs, and spike-and-slab estimation
operates on their \(-1\) entries; the first two are the depth-2
backbones used throughout, two anchor items per factor, and the third
specifies every primary loading:

\begin{itemize}
\tightlist
\item
  \textbf{Anchor-only (AO)}: the two anchor items of each group factor
  carry only their \(1\); no fixed zeros anywhere, so every non-anchor
  entry is \(-1\).
\item
  \textbf{Anchor-zero (AZ)}: the same two anchors, whose rows now carry
  fixed \(0\) on the other specified columns, \((1, 0, \dots, 0)\)-type
  rows; every other entry \(-1\).
\item
  \textbf{Full-primary (FP)}: every primary loading specified (\(1\)),
  every off-pattern entry \(-1\), complete pattern knowledge, but no
  hard zeros. This is the typical design of applied PCFA practice
  (\citeproc{ref-chen2021scale}{Chen et al., 2021};
  \citeproc{ref-chen2021partially}{Chen, 2021}), and it serves here only
  as the upper reference point of the preliminary study. Each name
  denotes a \emph{design}, not a model: an anchor-zero design serves
  oblique and bifactor fits alike.
\end{itemize}

AO and AZ therefore use the same two anchors per factor and differ only
in whether the anchors' cross-loadings are estimated under
regularization (AO) or fixed to zero (AZ). This \textbf{specification
depth}--the number of anchors per factor--is distinct from persistence
depth \(r\), the number of larger candidate solutions checked below.

\subsection{Distinguishability: two questions, one
signal}\label{sec-disting}

The phrase ``is the bifactor identified?'' conflates two questions that
the literature answers separately. Keeping them apart organizes
everything that follows. We give them their proper names.

\textbf{Identifiability is the estimation-side property}: \emph{given}
that the orthogonal bifactor model is true, are its parameters uniquely
determined by the covariance matrix, so that an estimator can recover
them? This is a property of the model and its design: how many items per
cluster, and which loadings are anchored. The question is settled
essentially completely by Fang et al.
(\citeproc{ref-fang2021identifiability}{2021}): for the standard
(orthogonal) bifactor used here, identifiability holds if and only if
every group factor has at least three items with nonzero group loadings
and at least three clusters carry nonzero general loadings (or two, with
a rank condition); for the extended model with correlated group factors,
their sufficient conditions count clusters whose general and group
loadings are linearly independent within the cluster, exactly our
within-cluster non-proportionality. When identifiability holds, anchored
estimation is routine. The simulations below show it remains routine
even in populations where the general factor is empirically redundant.

\textbf{Distinguishability is the population-side property}: is the
\emph{general factor} needed at all? That is, can the same covariance
matrix be reproduced exactly by a \(K\)-factor model, so that no
estimator, however good, could distinguish the two? This is a property
of the population loading pattern, not of any design or algorithm. It is
the Schmid--Leiman lineage's question
(\citeproc{ref-mansolf2017does}{Mansolf \& Reise, 2017};
\citeproc{ref-schmid1957development}{Schmid \& Leiman, 1957};
\citeproc{ref-waller2018direct}{Waller, 2018};
\citeproc{ref-yung1999relationship}{Yung et al., 1999}), and the one
that matters for model \emph{choice}. Identifiability asks whether the
parameters can be \emph{estimated}, given the structure.
Distinguishability asks whether the structure can be \emph{chosen},
given the population. No amount of data or estimation skill delivers
distinguishability at the reducible boundary, because there is nothing
in the covariance to deliver it with.

\subsubsection{Covariance equivalence on the reducible
set}\label{sec-prop1}

Two covariance-level statements organize distinguishability, and one
decomposition carries both. This subsection gives the first and Section
\ref{sec-thm1} the second. Write \(\mathbf{b}_{g|k}\) for the
restriction of the general column to cluster \(k\) (every cluster
carries nonzero group loadings), and let \(\Psi\) denote the diagonal
matrix of uniquenesses, so \(\Sigma = C + \Psi\). Within each cluster,
decompose \[
\mathbf{b}_{g|k} = \alpha_k \mathbf{u}_k + \mathbf{w}_k, \qquad \mathbf{u}_k = \mathbf{b}_s^k / \lVert \mathbf{b}_s^k \rVert, \quad \mathbf{w}_k \perp \mathbf{b}_s^k,
\] so that \(\alpha_k\) measures the general loadings' component along
the cluster's group loadings and the residual \(\mathbf{w}_k\) measures
the departure from proportionality. A cluster with
\(\mathbf{w}_k = \mathbf{0}\) has exactly proportional loadings and is
called \textbf{silent}; under the usual bifactor identifiability
conditions its parameters can remain identified within the chosen
bifactor class (with the additional rank qualification required in the
two-cluster case), but it contributes no direct within-cluster
nonproportionality signal of its own. Its presence can still change the
global rank budget and hence distinguishability in a mixed
configuration. Everything that follows is a statement about the residual
pattern \(\{\mathbf{w}_k\}\).

\textbf{Proposition 1 (reducibility implies equivalence).} If
\(\mathbf{w}_k = \mathbf{0}\) in every cluster, so that
\(\mathbf{b}_{g|k} = c_k\mathbf{b}_s^k\) with
\(c_k = \alpha_k/\lVert\mathbf{b}_s^k\rVert\), then \(C = A\Phi A'\)
exactly, with second-order loadings \(\gamma_k = c_k/\sqrt{1+c_k^2}\)
(\citeproc{ref-yung1999relationship}{Yung et al., 1999}). Equivalently,
the ratio defined above is \(b_{s,j}/b_{g,j}=1/c_k\) within cluster
\(k\). The same uniquenesses serve both representations, so \(\Sigma\)
admits an exact \(K\)-factor representation. A reducible bifactor is
indistinguishable from correlated factors at any sample size.

\subsubsection{A sufficient condition for
distinguishability}\label{sec-thm1}

The converse direction is more delicate than a rank count suggests. The
common component \(C\) has rank \(K^* = K+1\) whenever some cluster is
non-proportional, and any \(K\)-factor common part has rank at most
\(K\). But model equivalence is a statement about \(\Sigma\), not about
\(C\). A rival \(K\)-factor representation may assign different
uniquenesses: \(\Sigma = C + \Psi = \tilde{C} + \tilde{\Psi}\) with
\(\tilde{\Psi}\) diagonal forces only \(\tilde{C} = C + \Delta\) for
some diagonal \(\Delta\). The correct question is whether any diagonal
perturbation of \(C\) has rank \(K\), and diagonal freedom is not
innocent: the boundary analysis below exhibits configurations in which a
non-proportional cluster is traded away into the uniquenesses exactly.

\textbf{Theorem 1 (distinguishability).} Let \(\Sigma = C + \Psi\) arise
from an orthogonal bifactor structure with \(K \ge 2\) clusters, a
nonzero general loading on every item, at least three items per cluster,
and at least two nonzero group loadings per cluster. If
\(\mathbf{w}_k \ne \mathbf{0}\) in every cluster, then no \(K\)-factor
model reproduces \(\Sigma\) under any diagonal uniquenesses:
\(\operatorname{rank}(C + \Delta) \ge K + 1\) for every diagonal
\(\Delta\). The Appendix gives the proof.

The proof idea is short. Any rival differs from \(C\) by a diagonal
matrix, and \(C + \Delta = \mathbf{b}_g\mathbf{b}_g' + M\) with \(M\)
block diagonal over clusters. One algebraic fact does the work (Lemma A3
of Appendix A): with three or more items, a multiple of
\(\mathbf{b}_{g|k}\mathbf{b}_{g|k}'\) can differ from
\(\mathbf{b}_s^k\mathbf{b}_s^{k\prime}\) by a diagonal matrix only when
\(\mathbf{w}_k = \mathbf{0}\) or the group loading is a single-indicator
spike. Each cluster with \(\mathbf{w}_k \ne \mathbf{0}\) therefore
pushes one dimension of rank through its block, \(K\) such clusters
exhaust a rank-\(K\) budget, and the rival's column space then confines
every block's range to the span of its restricted general column, which
is the excluded proportional configuration once more. Notably, three
items per cluster suffice; no further pattern condition is required.

\subsubsection{The mixed boundary}\label{sec-boundary}

When some clusters are proportional, the theorem's hypothesis fails and
the rank budget acquires slack, and the \emph{pattern} of a
non-proportional cluster can then matter. Call cluster \(k\)
\textbf{resistant} if it has at least four items and its within-cluster
ratio vector \(\rho_i = b_{s,i}/b_{g,i}\) contains two disjoint unequal
pairs; equivalently, if \(\rho^k\) is not constant after deleting at
most one entry. The selected checks in Table~\ref{tbl-boundary}
reproduce near-zero residuals for analytically constructed exact rivals
when the lone non-proportional cluster has only three items or a
constant-except-one ratio pattern, while the tested resistant
configurations retain positive local minima. These examples illuminate
but do not locate a universal boundary: a formal characterization of the
mixed case remains open, and positive numerical minima are evidence
rather than proof of separation.

\subsubsection{Graded distinguishability and the reducibility
tiers}\label{sec-estimand}

Proposition 1 and Theorem 1 together frame the estimand:
distinguishability is graded at the \(\Sigma\) level. Define the
population distance to the \(K\)-factor class as \[
D_K(\Sigma) \;=\; \min_{\substack{\Lambda \in \mathbb{R}^{J \times K},\, \tilde\Psi \ge 0 \text{ diagonal}\\ \Lambda\Lambda' + \tilde\Psi \succ 0}} F_{\mathrm{ML}}\!\left(\Sigma,\; \Lambda\Lambda' + \tilde\Psi\right),
\] where
\(F_{\mathrm{ML}}(\Sigma, \Omega) = \operatorname{tr}(\Omega^{-1}\Sigma) - \log\lvert\Omega^{-1}\Sigma\rvert - J\)
is the maximum likelihood discrepancy. Proposition 1 says \(D_K = 0\) on
the reducible set. The displayed minimum is attained, and where Theorem
1 excludes an exact \(K\)-factor representation the attained value is
strictly positive rather than a nonattained zero infimum; Appendix A
gives the compactness argument. The resulting population
\textbf{necessity estimand} is: \[
H_0\colon D_K(\Sigma) = 0 \qquad \text{versus} \qquad H_1\colon D_K(\Sigma) > 0,
\] where \(H_0\) states that \(K\) common dimensions suffice for the
covariance, not that a higher-order or proportional bifactor
interpretation is false; on the reducible set the oblique, higher-order,
and proportional bifactor representations belong to the same
covariance-equivalence class, and preferring the \(K\)-dimensional
representation there is a parsimony decision, not the recovery of a
uniquely true class. Section \ref{sec-route} operationalizes this
question with selected anchored model classes. That comparison is
deliberately narrower than the unrestricted minimization defining
\(D_K\): it is evidence about whether an added general column improves
on the delivered anchored structure, not an exact sample estimator or
formal test of unrestricted \(D_K\). Substantive interpretation is a
third step, requiring interpretable loadings and robustness to residual
dependence. The magnitudes involved matter for everything downstream.
Numerically, for the moderate bifactor population P1 (Table
\ref{tbl-sim1pop}), \(D_4 = 1.251\) (an omitted group factor), while
\(D_5 = .031\) (the general dimension's residual gap) and \(D_6 = 0\):
the general dimension leaves a real but forty-fold smaller gap than a
missing group factor, which motivates the two-step division of labor
without asserting that the operational comparison equals \(D_K\).

Two geometric facts drive the intuition. Cross-cluster covariance blocks
have rank at most one under the bifactor and simple-structure oblique
models; cross-loadings can raise their rank in the general oblique
model. The bifactor's cross-cluster blocks admit a correlated-factor
representation, so those blocks alone do not establish the need for a
general factor. Distinction depends on reproducing them together with
the within-cluster covariance, and clusters with small
\(\lVert\mathbf{w}_k\rVert\) bring the models close to empirical
equivalence at realistic \(N\).

\textbf{Reducibility tiers.} Populations are organized by their residual
pattern: \textbf{Tier I} (theorem-covered distinguishability:
\(\mathbf{w}_k \ne \mathbf{0}\) in every cluster, none a
single-indicator spike), \textbf{Tier II} (the mixed boundary: at least
one silent and at least one non-proportional cluster), and \textbf{Tier
III} (reducible: \(\mathbf{w}_k = \mathbf{0}\) everywhere; exactly HO by
Proposition 1). Tier I includes near-proportional configurations: the
theorem still establishes separation, although \(D_K\) and empirical
power can be small. Tier II is not a single equivalence class. Selected
searches for P3 and P4 retain positive fitted minima (\(D_5 = .017\) and
\(.002\); Table~\ref{tbl-boundary}), providing numerical evidence of
separation, while other mixed patterns admit analytically constructed
exact \(K\)-factor rivals. Its status depends on the within-cluster
ratio patterns, and its empirical evidence can degrade as silent
clusters consume the rank budget. The asymptotic separation signal
scales with \(N \cdot D_K\), the sample-size-weighted distance to the
best \(K\)-factor model; finite-sample power also depends on the
estimation and decision procedure, sampling variability, and the local
null and nuisance geometry, not on an eigenvalue gap alone.

\subsubsection{Four questions, one bridge}\label{sec-fourq}

Everything that follows rests on keeping four questions apart.
\emph{Identifiability} asks whether a specified model's parameters can
be recovered; \emph{count selection} asks which candidate counts deserve
examination; \emph{structural stability} asks whether approximately the
same configuration persists in direct comparisons to declared larger
candidates; and \emph{distinguishability} asks whether
\(D_K(\Sigma) > 0\), so that an added dimension is required to reproduce
the covariance. The middle two are logically independent, since a
unanimously selected count can carry a structure that fails to persist
and neighboring candidates can share a persisting core. Step 2 poses the
last question conditionally on the delivered structure, as evidence
about the population property rather than an estimator of it. Nor do the
four collapse: P5 is identified yet covariance-equivalent to a
higher-order model, a stable structure can sit near the reducible
boundary, and an under-counted or locally dependent structure can
manufacture an apparently necessary general factor. Stable structure is
the bridge between estimation and the bifactor comparison, neither a
consequence of identifiability nor proof of distinguishability.

\subsection{Estimation and operational criteria}\label{sec-estimation}

All fits use the variational Bayes estimator of the PCFA framework,
implemented in the \textbf{vbpm} package
(\citeproc{ref-chen2026vbpm}{Chen \& Jin, 2026b};
\citeproc{ref-jin2025regularized}{Jin \& Chen, 2025a}), which spans the
confirmatory--exploratory continuum through regularization on the
unspecified entries (\citeproc{ref-chen2020pcirt}{Chen, 2020},
\citeproc{ref-chen2021partially}{2021},
\citeproc{ref-chen2022lawbl}{2022}, \citeproc{ref-chen2023fully}{2023};
\citeproc{ref-chen2021scale}{Chen et al., 2021};
\citeproc{ref-jinchen2025mimic}{Jin \& Chen, 2025b}). Residual
covariance is diagonal, and every unspecified loading receives a
posterior inclusion probability, hard-selected at \(\ge .5\). Four score
summaries have bounded scientific roles. ELBO- and BIC-gain paths supply
co-primary descriptive count evidence at the 20\% cut with sustain one;
the prospective and publication-reading chronology is specified below.
Hard variational BIC is the operational Step-2 criterion; \(\Delta\)ELBO
remains an algorithmic sensitivity because its cross-design
comparability is not established. The PEFA variational counterparts of
conventional SEM fit indices provide conservative diagnostic screens and
remain ancillary to the count and structural-delivery rules.
Package/function ownership, fitting controls, and software provenance
are documented in the replication archive README.

\subsection{The two-step rule and process}\label{sec-route}

Given data with suspected clusters and unknown structure, the process
separates two decisions that are confounded in common practice: how many
first-order factors the data support, and whether a general factor sits
above them. The deliverable is a stable structure when one exists, with
the count as an attribute of the structure it travels with. Uncertainty
can run upward (the selected structure may persist inside a larger one)
or downward (a smaller core may be the stable object), so the process
reports where the structure is stable rather than defending a single
number, and reporting that no stable structure was delivered is a
legitimate output when none is stable enough. Nothing in the process
consults the true or theoretical factor count.

\textbf{The two sweep modes, formally.} A factor-structure sweep
launches from a backbone \(Q_0\) (\(J \times K_0\) with \(K_0 \le K\);
two anchors per backbone factor) and adds \(K-K_0\) fully exploratory
(\(-1\)) columns to reach each candidate count in a fixed window
\([K_{\min}, K_{\max}]\), \(K_{\min} \ge K_0\), bracketing the counts
suggested by classical extraction methods
(\citeproc{ref-braeken2017empirical}{Braeken \& Assen, 2017};
\citeproc{ref-horn1965rationale}{Horn, 1965};
\citeproc{ref-kaiser1960application}{Kaiser, 1960}). The base candidate
\(K=K_0\) uses the backbone without padding. In \textbf{oblique mode},
candidate \(K\) fits the correlated-factors model with \(K\) columns and
free \(\Phi\). In \textbf{bifactor mode}, the same \(Q_0\) is used, a
fully specified general column is added internally, and all factors are
orthogonal; \(K\) still counts group factors, so candidate \(K\) carries
\(K^* = K + 1\) columns. The preliminary mode comparison recorded the
prespecified ELBO/BIC gain and raw-criterion readings in both modes. Its
result supports using the oblique sweep within the studied conditions;
it does not establish universal dominance.

The held-out stage of the major study used one AO oblique fit window and
a prospectively declared ELBO count profile over \(3{:}9\). For
publication, the complete fitted window \(W\) is read twice:
\(\widehat K_E(W)\) uses ELBO and \(\widehat K_B(W)\) uses
\(-\mathrm{BIC}\). For each path, the complete fitted window first
determines the largest positive gain and its last occurrence. Scanning
then begins after that transition and returns the first \(K\) whose gain
is strictly below 20\% of the maximum (sustain one). If the maximum is
not positive or no such crossing occurs, the reading is absent; there is
no boundary fallback. Gains after the first qualifying crossing do not
affect the reading. Their finite values form
\(C_{20}(W)=\{\widehat K_E(W),\widehat K_B(W)\}_{\mathrm{finite}}\);
whether one or both criteria support a delivered count is descriptive
only. The score orientation, crossing rule, and unusable-path conditions
are given in the online supplement. These readings identify a
descriptive count neighborhood, not the structural delivery decision.
Because the dual/full-window convention was defined after held-out
analysis, its L1/L2 partition is descriptive rather than prospectively
confirmed.

\textbf{Step 1: the PEFA sweep, with backbones and stability analysis.}
The sweep is a partially exploratory factor analysis (PEFA): a PCFA
design whose entries are mostly unspecified, evaluated over candidate
counts from an anchored backbone (\citeproc{ref-chen2023fully}{Chen,
2023}; \citeproc{ref-chenjin2026fit}{Chen \& Jin, 2026a}). Its inputs
are item-level data, a cluster hypothesis, and two anchor items per
suspected cluster, taken in instrument order unless substantive markers
are prespecified. The target is \emph{not} the common-factor rank of
\(\Sigma\): in an irreducible bifactor population no exact \(K\)-factor
representation exists, so a rank-consistent criterion may legitimately
select \(K + 1\) at large \(N\). What separates the steps is the
magnitude ordering of Section \ref{sec-estimand}, where an omitted group
factor produces block misfit an order of magnitude larger than the
general dimension's residual gap. Step 1 therefore targets the
\textbf{stable group-structure core}, the configuration of first-order
clusters that persists under declared direct wider-count comparisons,
with the count as its attribute; the gain rule identifies the
neighborhood, while structural persistence supplies the delivery
evidence. Both the anchor-only and anchor-zero backbones are plausible
designs in either sweep mode, and the simulation studies of Section
\ref{sec-sims} decide between them; scientific window declarations are
in the online supplement and fixed-window implementation details in the
replication archive README.

\textbf{The stability analysis and its decision rules.} For each direct
pair \(k<l\), the backbone columns are fixed by position and every
remaining source column is paired to the target with minimum
sign-aligned squared loading distance. The two decision metrics are
evaluated on those same assignments, and reuse of a target is recorded
as a collision that fails the affected edge.
Table~\ref{tbl-stability-terms} defines the reader-facing terms; exact
formulas, tie rules, missing-evidence precedence, and the descriptive
transition diagnostics are given in the online supplement.

\begin{longtable}[]{@{}
  >{\raggedright\arraybackslash}p{(\linewidth - 4\tabcolsep) * \real{0.2376}}
  >{\raggedright\arraybackslash}p{(\linewidth - 4\tabcolsep) * \real{0.4653}}
  >{\raggedright\arraybackslash}p{(\linewidth - 4\tabcolsep) * \real{0.2970}}@{}}
\caption{Reader-facing definitions for the Step-1 stability analysis. AO
and AZ are defined in Section \ref{sec-notation}. Full measurement
definitions and optional PIP diagnostics are in the online
supplement.}\label{tbl-stability-terms}\tabularnewline
\toprule\noalign{}
\begin{minipage}[b]{\linewidth}\raggedright
Term
\end{minipage} & \begin{minipage}[b]{\linewidth}\raggedright
Definition and direction
\end{minipage} & \begin{minipage}[b]{\linewidth}\raggedright
Role
\end{minipage} \\
\midrule\noalign{}
\endfirsthead
\toprule\noalign{}
\begin{minipage}[b]{\linewidth}\raggedright
Term
\end{minipage} & \begin{minipage}[b]{\linewidth}\raggedright
Definition and direction
\end{minipage} & \begin{minipage}[b]{\linewidth}\raggedright
Role
\end{minipage} \\
\midrule\noalign{}
\endhead
\bottomrule\noalign{}
\endlastfoot
Minimum congruence, \(\phi_{\min}\) & the smallest absolute Tucker
congruence across the matched source columns; larger means the weakest
matched shape is more similar & reference persistence metric for the
three profiles \\
RMSD and maximum RMSD & a column root-mean-square difference (RMSD)
summarizes its sign-aligned loading differences; pooled RMSD aggregates
all matched cells, while \(\mathrm{RMSD}_{\max}\) is the largest column
RMSD, so smaller means greater worst-column similarity &
\(\mathrm{RMSD}_{\max}\) is the alternative persistence metric; pooled
RMSD is descriptive \\
ARI & adjusted Rand index for agreement between the two solutions'
dominant-factor item partitions, corrected for chance & descriptive
only; it does not determine delivery \\
SSL and unmatched SSL & SSL is a column's sum of squared loadings;
unmatched SSL is the largest SSL among unused target columns &
descriptive surplus-strength evidence only \\
Persistence depth, \(r\) & the number of larger candidates checked
directly: source \(K\) must pass every edge \(K\to K+1,\ldots,K+r\) &
defines how locally or deeply stability is required; comparisons are
direct, not chained \\
Persistence count, \(K_p\) & the highest persistent source after every
higher source in the declared set resolves nonpersistent &
profile-selected in-window count; neither true \(K\) nor necessarily
either \(\widehat K_E\) or \(\widehat K_B\) \\
Count-evidence set, \(C_{20}(W)\) & the finite ELBO- and BIC-gain
readings at the 20\% cut, sustain one, over the complete fitted window
\(W\) & descriptive count evidence; one-versus-both support does not
change delivery \\
\end{longtable}

A \textbf{persistence profile} declares a metric, cutoff, fixed source
set, and depth \(r\). The \textbf{source set} is the declared subset of
the sweep window from which persistence is read, and it is necessarily
narrower than that window: a source \(K\) evaluated at depth \(r\)
requires the sweep to reach \(K+r\), so the top \(r\) counts exist only
to supply forward comparisons and can never themselves be sources.
Declaring the set in advance prevents searching upward for a persistent
count after the fact. A \(K_p\) at the highest declared source is
potentially ceiling-limited, but is demonstrably truncated only if a
prespecified or sensitivity extension delivers a higher source. A source
persists only when every required direct edge passes. A higher
unavailable source makes \(K_p\) unresolved, whereas an unavailable
lower source does not overturn an already resolved higher persistent
source. If every source resolves nonpersistent, \(K_p\) is validly
absent. Sections \ref{sec-study3} and \ref{sec-transition} show that an
accurate count reading can coexist with a different persistence reading.
The layer is then given by Table~\ref{tbl-layers}:

\begin{longtable}[]{@{}
  >{\raggedright\arraybackslash}p{(\linewidth - 6\tabcolsep) * \real{0.0822}}
  >{\raggedright\arraybackslash}p{(\linewidth - 6\tabcolsep) * \real{0.2603}}
  >{\raggedright\arraybackslash}p{(\linewidth - 6\tabcolsep) * \real{0.2603}}
  >{\raggedright\arraybackslash}p{(\linewidth - 6\tabcolsep) * \real{0.3973}}@{}}
\caption{The delivery decision: two axes of evidence, three substantive
layers, and an unclassified computation/evidence
state.}\label{tbl-layers}\tabularnewline
\toprule\noalign{}
\begin{minipage}[b]{\linewidth}\raggedright
Layer
\end{minipage} & \begin{minipage}[b]{\linewidth}\raggedright
Count evidence
\end{minipage} & \begin{minipage}[b]{\linewidth}\raggedright
Structural evidence
\end{minipage} & \begin{minipage}[b]{\linewidth}\raggedright
Reporting outcome
\end{minipage} \\
\midrule\noalign{}
\endfirsthead
\toprule\noalign{}
\begin{minipage}[b]{\linewidth}\raggedright
Layer
\end{minipage} & \begin{minipage}[b]{\linewidth}\raggedright
Count evidence
\end{minipage} & \begin{minipage}[b]{\linewidth}\raggedright
Structural evidence
\end{minipage} & \begin{minipage}[b]{\linewidth}\raggedright
Reporting outcome
\end{minipage} \\
\midrule\noalign{}
\endhead
\bottomrule\noalign{}
\endlastfoot
L1 & \(K_p\in C_{20}(W)\) & persistent & deliver the structure; record
whether ELBO, BIC, or both supplied support \\
L2 & \(K_p\notin C_{20}(W)\) and at least one count path is usable &
persistent & deliver the persistent structure; carry the count
uncertainty into step 2 \\
L3 & count paths may be usable or unusable & all sources resolved
nonpersistent & report that no stable structure was delivered \\
unclassified & any count evidence & persistence unresolved, or
persistent \(K_p\) when both count paths are unusable & report the
computation/evidence limitation; do not reinterpret it as L3 \\
\end{longtable}

L1 requires persistent finite \(K_p\) to equal at least one finite
member of \(C_{20}(W)\). L2 requires a persistent \(K_p\) outside that
set and at least one usable path; the usable path may return a different
finite count or validly return no count. L3 requires every source to
resolve nonpersistent and \(K_p\) to be validly absent, irrespective of
count-path usability. Unresolved \(K_p\), or persistent \(K_p\) when
both count paths are unusable, is unclassified. Missing or malformed
persistence evidence never becomes L3. Criterion support labels
(\texttt{both}, \texttt{ELBO\ only}, \texttt{BIC\ only}, or
\texttt{none}) are descriptive attributes, not sublayers. ARI, pooled
RMSD, SSL, the count-to-persistence gap introduced below, and the
remaining diagnostics do not create another delivery branch. The L3
report, \textbf{no stable structure was delivered} (non-delivery), means
the declared evidence did not deliver a first-order structure stable
enough to support the conditional bifactor comparison; it is procedural,
not expert judgment. Substantive expertise guides anchors, interprets
competing solutions, and motivates follow-up; a structure can be imposed
after non-delivery as a \emph{substantively specified sensitivity
analysis}.

\textbf{Step 2: the operational model comparison.} Only an L1/L2
delivery proposes Step 2, and then \(K_{\mathrm{step2}}=K_p\). The
delivered AO loading matrix supplies two markers per added cluster while
the inherited backbone is retained exactly; the resulting oblique and
bifactor arms share one \(J\times K_{\mathrm{step2}}\) group-design
matrix, with the bifactor general column added internally. Identical
dataset/AO/\(K_{\mathrm{step2}}\) proposals from different profiles are
fitted once and mapped back to every originating profile. Hard BIC
compares the two arms. For empirical reporting the amendment also fixed,
before any held-out artifact existed, a pre-held-out reporting band of
\(\lvert\Delta\mathrm{BIC}\rvert \le 2\), within which the arms show no
practically meaningful BIC separation while the signed value is still
reported. Parameter evaluation of a consistently preferred
representation reports loadings, factor correlations, and general and
group effect sizes (\citeproc{ref-rodriguez2016evaluating}{Rodriguez et
al., 2016}; \citeproc{ref-zhang2024accommodating}{Zhang \& Chen, 2024}),
conditional on external evidence.

\section{Simulation studies}\label{sec-sims}

The simulation program consists of two studies, each with two explicitly
named stages. The labels A and B distinguish the preliminary-study
stages from both the procedure's Step 1 and Step 2 and the simulation
populations P1--P6. Table~\ref{tbl-sim-overview} summarizes the two
studies and their internal stages.

\begin{longtable}[]{@{}
  >{\raggedright\arraybackslash}p{(\linewidth - 6\tabcolsep) * \real{0.2500}}
  >{\raggedright\arraybackslash}p{(\linewidth - 6\tabcolsep) * \real{0.2500}}
  >{\raggedright\arraybackslash}p{(\linewidth - 6\tabcolsep) * \real{0.2500}}
  >{\raggedright\arraybackslash}p{(\linewidth - 6\tabcolsep) * \real{0.2500}}@{}}
\caption{The two simulation studies and their internal
stages.}\label{tbl-sim-overview}\tabularnewline
\toprule\noalign{}
\begin{minipage}[b]{\linewidth}\raggedright
Study
\end{minipage} & \begin{minipage}[b]{\linewidth}\raggedright
Stage
\end{minipage} & \begin{minipage}[b]{\linewidth}\raggedright
Scientific question
\end{minipage} & \begin{minipage}[b]{\linewidth}\raggedright
Evidentiary role
\end{minipage} \\
\midrule\noalign{}
\endfirsthead
\toprule\noalign{}
\begin{minipage}[b]{\linewidth}\raggedright
Study
\end{minipage} & \begin{minipage}[b]{\linewidth}\raggedright
Stage
\end{minipage} & \begin{minipage}[b]{\linewidth}\raggedright
Scientific question
\end{minipage} & \begin{minipage}[b]{\linewidth}\raggedright
Evidentiary role
\end{minipage} \\
\midrule\noalign{}
\endhead
\bottomrule\noalign{}
\endlastfoot
Preliminary & A: generating structure supplied & how well do the
anchored designs estimate the structure, and how does the conditional
model comparison behave at the generating count? & component evidence;
motivates rather than validates the final process \\
Preliminary & B: structure recovered through sweeps & which sweep mode
counts usefully, how does anchoring affect the count, and what do over-
and under-counts do to the retained structure? & tentative design
findings carried into development \\
Major & Developmental & which backbone, metric, cutoff, and persistence
depth should define the operating family? & develops and freezes the
three profiles and Step-2 rules \\
Major & Confirmatory (held out) & how do the frozen profiles and
conditional comparison perform on fresh replications from the same
generators? & scenario-bound confirmation without reselection \\
\end{longtable}

\subsection{Preliminary simulation study: Stages A and
B}\label{sec-prelim}

Before the process can be evaluated end to end, four design questions
must be examined: how deeply to anchor, which sweep mode carries the
count, whether the backbone's anchoring changes that count, and what an
over- or under-count does to the recovered structure. The preliminary
study uses six fixed populations to provide component evidence within
those conditions; Appendix B and online- supplement Tables S2--S5 report
both stages in full.

\textbf{Preliminary Stage A} supplies Preliminary Findings 1--2. AO
trails AZ slightly and both trail full-primary specification only where
estimation is already hard-- weak loadings at small \(N\) and the
reducible population--with omissions rather than false discoveries
driving the gaps (Preliminary Finding 1). At the generating count, the
conditional comparison has favorable operating characteristics in both
directions (Preliminary Finding 2). These are tentative component
findings at a supplied structure, not end-to-end evidence about
persistence.

\textbf{Preliminary Stage B} supplies Preliminary Findings 3--6. The
oblique sweep counts more reliably than the bifactor sweep under either
backbone because surviving columns' free entries can absorb an omitted
group factor's covariance block (Preliminary Finding 3). AO and AZ count
paths are close in this preliminary task, while an apparent anchoring
advantage is traced to an accidentally over-restricted backbone
(Preliminary Finding 4). Mild oblique over-counts retain the major core
beside a thin surplus column in the studied populations (Preliminary
Finding 5), whereas bifactor-mode under-counts remove a group direction
and can manufacture a general factor (Preliminary Finding 6).

These preliminary findings motivate using an oblique sweep in the major
study; that sweep mode is not re-tested there. They do not settle the
backbone or the persistence rule. In particular, preliminary
near-agreement of AO and AZ count paths does not imply equal structural
persistence: the major developmental stage asks that different question.
The phi-versus-RMSD comparison and the three-profile family are also new
major-study questions, not conclusions of the preliminary study.

\subsection{Major simulation study: developmental and confirmatory
stages}\label{sec-study3}

The major study runs the full two-step process of Section
\ref{sec-route} under model uncertainty. Its central safeguard is the
separation of development from confirmation by a prospectively frozen
protocol rather than a random split. A completed \textbf{developmental
stage} (replication IDs 1--200) crossed both backbone specifications and
both count criteria over a grid of 25 persistence candidates at three
depths and developed the operating rules against generating truth. A
postdevelopment amendment froze those rules, the three reference
profiles, one practical default, the delivery-matched comparators, and
every Step-2 construction and reporting rule before any confirmatory
artifact existed. A \textbf{held-out confirmatory stage} (IDs 201--400)
then estimated the frozen rules once on fresh replications. Here
\emph{confirmatory} means a prospective, scenario-bound held-out
evaluation: it is not external validation and cannot promote, delete,
retune, or reselect a rule. Because AZ was not fitted at this stage, it
confirms AO's frozen operating characteristics rather than re-testing
the developmental AO-over-AZ contrast.

\subsubsection{Design}\label{sec-s3-design}

Table~\ref{tbl-s3pop} lists the seven populations: the three structural
bases of the preliminary study crossed with interference, plus one
weak-core base. The \emph{doublet} adds residual covariance \(c\)
between the two non-anchor items of cluster 1 (loadings \(.6\) and
\(.3\); positive definiteness requires \(c < .251\)); a calibration
pilot located the regimes, with \(c = .22\) genuinely borderline and
\(c = .20\) the strongest fully absorbed strength. The \emph{minor
factor} is orthogonal, loads on every item with magnitudes drawn from
\((.09, .11)\) and random signs following Auerswald and Moshagen
(\citeproc{ref-auerswald2019determine}{2019}), and is redrawn each
replication. \textbf{P2} is a weak-core bifactor whose general loadings
are \(.70\) but whose group loadings alternate \(.20\) and \(.40\), so
that the weakest group directions sit at or below detectability at these
sample sizes; its five group columns remain the Step-1 truth, and it is
never declared a population without structure. The adversarial
population pairs the higher-order base with the strongest fully absorbed
dependence: whatever the doublet contributes flows into the Step-2
comparison, and Theorem 1's diagonal-uniqueness assumption fails exactly
as residual dependence makes it fail.

\begin{longtable}[]{@{}
  >{\raggedright\arraybackslash}p{(\linewidth - 6\tabcolsep) * \real{0.2500}}
  >{\raggedright\arraybackslash}p{(\linewidth - 6\tabcolsep) * \real{0.2500}}
  >{\raggedright\arraybackslash}p{(\linewidth - 6\tabcolsep) * \real{0.2500}}
  >{\raggedright\arraybackslash}p{(\linewidth - 6\tabcolsep) * \real{0.2500}}@{}}
\caption{The seven populations of the major study: the three structural
bases crossed with interference, plus the weak-core
base.}\label{tbl-s3pop}\tabularnewline
\toprule\noalign{}
\begin{minipage}[b]{\linewidth}\raggedright
Population
\end{minipage} & \begin{minipage}[b]{\linewidth}\raggedright
Base (generating model)
\end{minipage} & \begin{minipage}[b]{\linewidth}\raggedright
Interference
\end{minipage} & \begin{minipage}[b]{\linewidth}\raggedright
Role
\end{minipage} \\
\midrule\noalign{}
\endfirsthead
\toprule\noalign{}
\begin{minipage}[b]{\linewidth}\raggedright
Population
\end{minipage} & \begin{minipage}[b]{\linewidth}\raggedright
Base (generating model)
\end{minipage} & \begin{minipage}[b]{\linewidth}\raggedright
Interference
\end{minipage} & \begin{minipage}[b]{\linewidth}\raggedright
Role
\end{minipage} \\
\midrule\noalign{}
\endhead
\bottomrule\noalign{}
\endlastfoot
P1 & bifactor & none & clean reference; general factor present \\
P1 + minor & bifactor & minor factor & diffuse interference \\
P1 + doublet & bifactor & doublet, \(c = .22\) & the stability
boundary \\
P2 & bifactor, weak core & none & stable-yet-wrong hazard \\
P5 & higher-order & none & clean reducible; no separable general
factor \\
P5 + doublet & higher-order & doublet, \(c = .20\) & absorbed
dependence; adversarial for step 2 \\
P6 & oblique & none & clean; no general factor; true cross-loadings \\
\end{longtable}

{\footnotesize\emph{Note.} Bases are Preliminary Stage A's populations
P1, P5, and P6, unchanged; population P2 shares P1's general loadings
with group pattern \(.20/.40\) and correspondingly adjusted
uniquenesses. The doublet sits on the two non-anchor items of cluster 1.
The grid crosses the seven populations with
\(N \in \{500, 1000, 2000\}\) at 200 replications per stage cell.}

Per developmental replication, oblique sweeps ran from both backbones
over the window \([3, 9]\) with ELBO and BIC gain paths. The stability
analysis crossed the full minimum-congruence and maximum-column-RMSD
grids at depths one through three (online supplement). The developmental
stage froze a \textbf{family of three reference perspectives}
(Table~\ref{tbl-profiles}) rather than a single operating point:

\begin{longtable}[]{@{}
  >{\raggedright\arraybackslash}p{(\linewidth - 6\tabcolsep) * \real{0.2500}}
  >{\raggedright\arraybackslash}p{(\linewidth - 6\tabcolsep) * \real{0.2500}}
  >{\raggedright\arraybackslash}p{(\linewidth - 6\tabcolsep) * \real{0.2500}}
  >{\raggedright\arraybackslash}p{(\linewidth - 6\tabcolsep) * \real{0.2500}}@{}}
\caption{The three reference profiles developed before held-out
confirmation. The RMSD companions are descriptive comparisons and never
propose Step 2.}\label{tbl-profiles}\tabularnewline
\toprule\noalign{}
\begin{minipage}[b]{\linewidth}\raggedright
Profile
\end{minipage} & \begin{minipage}[b]{\linewidth}\raggedright
Persistence requirement
\end{minipage} & \begin{minipage}[b]{\linewidth}\raggedright
Perspective
\end{minipage} & \begin{minipage}[b]{\linewidth}\raggedright
Delivery-matched RMSD companion
\end{minipage} \\
\midrule\noalign{}
\endfirsthead
\toprule\noalign{}
\begin{minipage}[b]{\linewidth}\raggedright
Profile
\end{minipage} & \begin{minipage}[b]{\linewidth}\raggedright
Persistence requirement
\end{minipage} & \begin{minipage}[b]{\linewidth}\raggedright
Perspective
\end{minipage} & \begin{minipage}[b]{\linewidth}\raggedright
Delivery-matched RMSD companion
\end{minipage} \\
\midrule\noalign{}
\endhead
\bottomrule\noalign{}
\endlastfoot
\texttt{.85/r1} & \(\phi_{\min}\ge .85\) for \(K\to K+1\) & stringent,
shallow/local sensitivity & \texttt{.18/r1} \\
\texttt{.80/r2} & \(\phi_{\min}\ge .80\) for both \(K\to K+1\) and
\(K\to K+2\) & middle-depth practical default & \texttt{.20/r2} \\
\texttt{.70/r3} & \(\phi_{\min}\ge .70\) for all three direct edges
through \(K+3\) & deeper, more tolerant-per-edge sensitivity &
\texttt{.24/r3} \\
\end{longtable}

{\footnotesize\emph{Note.} All three profiles use the fixed source set
\(3{:}6\) in both stages of the major study. Each source is checked
directly against the next \(r\) fitted counts.}

The family is non-nested: increasing \(r\) tightens the number of
required direct comparisons while lowering the congruence cutoff relaxes
each edge. \texttt{.80/r2} was declared the single provisional practical
default on developmental grounds. After the formal \(r=1{:}3\) grid was
closed, a nonselecting, truth-centered source-5 retention-yield
diagnostic at fixed \texttt{.70} declined at \(r=4\). It was exploratory
rather than a full operational \(K_p\) comparison, because the existing
window could not evaluate every source at that depth, and it selected no
profile. The amendment prospectively widened the confirmatory fit window
to \([3,10]\) so that \(r=1{:}4\) could be compared over the same
sources.

The confirmatory stage fits AO only (the developmental backbone finding
below) and one oblique sweep per dataset over \([3,10]\). Its
prospective ELBO count path used the \([3,9]\) core, with \(K=10\)
reserved for fourth-depth and source-7 diagnostics; the publication
rereading applies both 20\%/sustain-one paths to the complete fitted
window. This requires no refitting and changes neither \(K_p\),
delivery, nor Step 2. All 4,200 held-out datasets fitted without
failure, and every summary below was independently reproduced from the
raw evidence.

\subsubsection{Developmental stage: Findings 1--4}\label{sec-s3-dev}

Four major findings carry into the freeze. Table~\ref{tbl-devevidence}
collects the evidence behind the first three, all on the six-population
calibration domain.

\begin{longtable}[]{@{}
  >{\raggedright\arraybackslash}p{(\linewidth - 6\tabcolsep) * \real{0.2143}}
  >{\raggedright\arraybackslash}p{(\linewidth - 6\tabcolsep) * \real{0.2143}}
  >{\raggedleft\arraybackslash}p{(\linewidth - 6\tabcolsep) * \real{0.2857}}
  >{\raggedleft\arraybackslash}p{(\linewidth - 6\tabcolsep) * \real{0.2857}}@{}}
\caption{Developmental evidence behind the frozen rules: backbone,
metric, and depth comparisons against generating
truth.}\label{tbl-devevidence}\tabularnewline
\toprule\noalign{}
\begin{minipage}[b]{\linewidth}\raggedright
Comparison
\end{minipage} & \begin{minipage}[b]{\linewidth}\raggedright
Evidence
\end{minipage} & \begin{minipage}[b]{\linewidth}\raggedleft
AO
\end{minipage} & \begin{minipage}[b]{\linewidth}\raggedleft
AZ
\end{minipage} \\
\midrule\noalign{}
\endfirsthead
\toprule\noalign{}
\begin{minipage}[b]{\linewidth}\raggedright
Comparison
\end{minipage} & \begin{minipage}[b]{\linewidth}\raggedright
Evidence
\end{minipage} & \begin{minipage}[b]{\linewidth}\raggedleft
AO
\end{minipage} & \begin{minipage}[b]{\linewidth}\raggedleft
AZ
\end{minipage} \\
\midrule\noalign{}
\endhead
\bottomrule\noalign{}
\endlastfoot
Backbone & best major yield, depth 1 / 2 / 3 & .835 / .845 / .847 & .329
/ .327 / .322 \\
Backbone & delivered-but-wrong fraction at the three frozen rows & --- &
.577 / .496 / .436 \\
Backbone & share of those errors in bifactor populations & --- & 98.4\%
/ 98.1\% / 98.1\% \\
Metric & best major yield, minimum congruence, depth 1 / 2 / 3 & .835 /
.845 / .847 & --- \\
Metric & best major yield, maximum-column RMSD, depth 1 / 2 / 3 & .816 /
.803 / .784 & --- \\
Depth & delivery at fixed .70, depth 1 / 2 / 3 & .9992 / .9950 / .9872 &
--- \\
Depth & recovery precision at fixed .70, depth 1 / 2 / 3 & .8338 / .8498
/ .8604 & --- \\
\end{longtable}

{\footnotesize\emph{Note.} Major yield is the proportion of all attempts
that both deliver and retain the five generating directions. The
delivered-but-wrong fraction is conditional on delivery at the named
row. Dashes mark quantities outside a comparison, not missing
computation.}

\textbf{Finding 1: the anchor-only backbone outperforms anchor-zero for
structural persistence.} The best major yield is more than 50 percentage
points higher under anchor-only at every depth
(Table~\ref{tbl-devevidence}), and the gap is not a delivery artifact:
at the three frozen rows a large minority of anchor-zero deliveries are
stable yet wrong, almost all of them in the bifactor populations and
with median matched congruence near zero, while anchor-zero precision in
the populations without a separable general dimension stays between
\(.97\) and \(.99\). This qualifies Preliminary Finding 4 in a
task-specific way, because similar count paths do not imply similar
structural persistence. A plausible interpretation consistent with the
theory is that anchored zeros close an outlet through which an
irreducible general dimension can otherwise be absorbed at
\(\widehat K + 1\), so the larger solution reorganizes its retained
columns instead, and the populations without such a dimension are
unaffected. Anchor-only is therefore frozen for both steps, and the
finding remains developmental because anchor-zero is not refitted in
confirmation.

\textbf{Finding 2: minimum matched congruence outperforms the
maximum-column RMSD at matched depth.} The ordering holds at all three
depths and widens as depth grows. It is conditional on the shared
minimum-SSE correspondence and the congruence-defined recovery score
rather than an intrinsic ranking of the two indices, and the
delivery-matched RMSD companions carry the comparison into confirmation.

\textbf{Finding 3: persistence depth helps through three direct
comparisons.} Holding the cutoff at \(.70\), delivery falls slowly while
recovery precision rises. The clear gain is from one required comparison
to two; the third mainly exchanges a little delivery for precision,
which is why the family spans depths one through three rather than
stopping at one.

\textbf{Finding 4: three useful, non-nested profiles support one
provisional default.} None contains another, since increasing depth
tightens a policy while lowering the cutoff relaxes it. Their
developmental operating characteristics form the left-hand entries of
Table~\ref{tbl-transport}. \texttt{.80/r2} was modestly favorable in the
difficult populations and occupies the middle depth, which is the
substantive ground for the default; the shallower and deeper profiles
retain distinct sensitivity roles.

\subsubsection{Confirmatory stage: Findings 5 and 6}\label{sec-s3-val}

All held-out fits completed, and Table~\ref{tbl-transport} pairs each
developmental value with its confirmatory estimate on the shared
six-population domain. The exact fit and profile censuses are reported
in the online supplement.

\begin{longtable}[]{@{}lrrr@{}}
\caption{Development-to-confirmation reproduction of the frozen
reference family: delivery, recovery precision, and recovery yield on
the six-population domain (entries are development /
confirmation).}\label{tbl-transport}\tabularnewline
\toprule\noalign{}
Reference profile & Delivery & Recovery precision & Recovery yield \\
\midrule\noalign{}
\endfirsthead
\toprule\noalign{}
Reference profile & Delivery & Recovery precision & Recovery yield \\
\midrule\noalign{}
\endhead
\bottomrule\noalign{}
\endlastfoot
\texttt{.85/r1} & .9914 / .9897 & .8451 / .8504 & .8378 / .8417 \\
\texttt{.80/r2} & .9789 / .9744 & .8610 / .8717 & .8428 / .8494 \\
\texttt{.70/r3} & .9872 / .9822 & .8604 / .8640 & .8494 / .8486 \\
\end{longtable}

{\footnotesize\emph{Note.} Six-population domain (the adversarial
population reported separately); 3,600 attempts per stage and profile.
Recovery credits the class applicable to the delivered count: a correct
lower core below five, full recovery at five, five-major retention
above. Exact numerators, denominators, and pointwise 95\% Wilson
intervals are in the online supplement.}

\textbf{Finding 5: the frozen family closely reproduces on fresh
replications from the same generators.} Every confirmatory precision
sits within about one point of its developmental value and every
delivery rate within half a point (Table~\ref{tbl-transport}), so the
three perspectives reproduce as a family rather than one surviving at
the expense of the others. The provisional \texttt{.80/r2} default again
posts the highest precision, which retains it without post hoc
reselection.

\begin{longtable}[]{@{}lrrr@{}}
\caption{Per-population recovery precision at the three reference rows:
development / confirmation.}\label{tbl-perpop}\tabularnewline
\toprule\noalign{}
Population & \texttt{.85/r1} & \texttt{.80/r2} & \texttt{.70/r3} \\
\midrule\noalign{}
\endfirsthead
\toprule\noalign{}
Population & \texttt{.85/r1} & \texttt{.80/r2} & \texttt{.70/r3} \\
\midrule\noalign{}
\endhead
\bottomrule\noalign{}
\endlastfoot
P1 & .957 / .950 & .972 / .960 & .973 / .958 \\
P1 + minor & .938 / .945 & .943 / .946 & .945 / .958 \\
P1 + doublet & .870 / .870 & .877 / .884 & .865 / .858 \\
P2 & .349 / .391 & .407 / .455 & .388 / .409 \\
P5 & .985 / .971 & .983 / .986 & .998 / .991 \\
P6 & .966 / .968 & .963 / .973 & .975 / .988 \\
\end{longtable}

{\footnotesize\emph{Note.} Pooled over \(N\); the out-of-domain
adversarial population reached \(.990\) at the default on held-out data.
Development values are the frozen calibration record; confirmation
values are the held-out estimates.}

\textbf{Finding 6: persistence delivers a reproducible in-window
reading, not necessarily a correct structure.} The per-population
anatomy reproduces with the aggregate (Table~\ref{tbl-perpop}), and it
is dominated by one population. P2, whose weakest group directions sit
at or below detectability, delivers at a high rate while its modal
delivered count is six rather than five (Table~\ref{tbl-p2anatomy}): the
weakest directions are simply absent from a stable six-factor solution,
the same datasets fail at every reference cutoff with the same delivered
count, and although the failure rate falls as \(N\) grows it remains the
largest of any population. Because the failing deliveries are themselves
stable, no cutoff on the developed grid repairs them, and recovery must
be reported per population rather than pooled.

\begin{longtable}[]{@{}lrl@{}}
\caption{The weak-core hazard at the default profile: delivered counts
and recovery outcomes for the 600 P2
attempts.}\label{tbl-p2anatomy}\tabularnewline
\toprule\noalign{}
Delivered count & Datasets & Recovery outcome \\
\midrule\noalign{}
\endfirsthead
\toprule\noalign{}
Delivered count & Datasets & Recovery outcome \\
\midrule\noalign{}
\endhead
\bottomrule\noalign{}
\endlastfoot
six (modal) & 307 & 276 fail five-major retention, 31 retain \\
five & 231 & 220 full recovery, 11 fail \\
three or four & 14 & all fail lower-core recovery \\
not delivered & 48 & --- \\
\end{longtable}

Two further regularities sharpen the finding; the online supplement
tabulates both (Table S14). First, the prospectively frozen ELBO count
path and the persistence reading separate cleanly: that path selects the
generating count in almost every dataset, its few overcounts fall in the
borderline doublet population and its few undercounts in P2 at the
smallest sample size. Delivered counts of six therefore arise from
persistence rather than from that count path in every population; what
is specific to P2 is that almost all of its six-deliveries fail
recovery, whereas elsewhere they retain the generating directions.
Second, the three profiles agree on whether a structure exists far more
often than on its exact count, and their count disagreements almost
always span a single adjacent count, with the shallow reading keeping
that boundary dimension and the deep reading pruning it. Precision also
moves in opposite directions across \(N\) by population, rising where
the core is weak and falling where it is clean, so the pooled rate is a
mixture of opposing trends; the developmental record shows the same
crossing.

One further regularity, computed after the freeze, corroborates how
locally the count and persistence evidence disagree. For finite \(K_p\)
and nonempty \(C_{20}(W)\) define \[
g_{\mathrm{CP}}(K_p;W)=\min_{k\in C_{20}(W)}|k-K_p|.
\] The quantity is undefined when \(K_p\) is absent or unresolved, or
when neither count path returns a finite value. Within finite delivered
readings, zero is algebraically equivalent to L1, so only a positive L2
gap adds descriptive evidence. Held-out L2 disagreements are almost
always a single count and never wider than two, and the developmental
record under the anchor-only backbone agrees; the retired anchor-zero
design supplies the only wider readings. Direction is more informative
than width in the populated cells. Against the generating truth, a
\(K_p\) above the count evidence corresponds to over-extraction in
99.7\% of the held-out reference-profile readings and the core usually
survives, whereas a \(K_p\) below it usually under-extracts (79.0\% at
gap one and every case at gap two) and usually fails recovery, the same
asymmetry Preliminary Findings 5 and 6 establish for the sweep itself.
The census behind these statements, the effect of the dual-criterion
convention on the L1/L2 partition, and the recovery rates by direction
are in the online supplement (Tables S27 and S28). These are post hoc
corroborations of locality under the studied conditions: neither the
count convention nor the gap selected a profile, changed \(K_p\) or
delivery, or altered any Step-2 proposal or fit.

Confirmation also re-reads Findings 2 and 3 on fresh data
(Table~\ref{tbl-confdiag}). The delivery-matched RMSD companions stay
behind the congruence rows at matched depth, confirming the direction of
the developmental metric preference while keeping the margins modest.
Along the fixed-cutoff depth sequence the gain again concentrates
through the third comparison, and the prospectively declared fourth
comparison adds no precision while costing delivery, so \(r=4\) remains
a diagnostic rather than a fourth profile.

\begin{longtable}[]{@{}lrr@{}}
\caption{Confirmatory readings of the metric and depth comparisons: the
congruence rows, their delivery-matched RMSD companions, and the
fixed-cutoff depth sequence, all on the six-population domain of
Table~\ref{tbl-transport}.}\label{tbl-confdiag}\tabularnewline
\toprule\noalign{}
Reading & Delivery & Recovery precision \\
\midrule\noalign{}
\endfirsthead
\toprule\noalign{}
Reading & Delivery & Recovery precision \\
\midrule\noalign{}
\endhead
\bottomrule\noalign{}
\endlastfoot
minimum congruence, depth 1 & .9897 & .8504 \\
minimum congruence, depth 2 & .9744 & .8717 \\
minimum congruence, depth 3 & .9822 & .8640 \\
delivery-matched RMSD, depth 1 & .9886 & .8328 \\
delivery-matched RMSD, depth 2 & .9792 & .8122 \\
delivery-matched RMSD, depth 3 & .9858 & .7929 \\
fixed .70, depth 1 & .9989 & .8420 \\
fixed .70, depth 2 & .9911 & .8545 \\
fixed .70, depth 3 & .9822 & .8640 \\
fixed .70, depth 4 & .9642 & .8574 \\
\end{longtable}

\subsubsection{The conditional comparison: Finding
7}\label{sec-s3-step2}

\textbf{Finding 7: conditional on a correctly recovered structure the
two-arm comparison is near-deterministic in both directions, and Step 1
is the system bottleneck.} Step 2 covered the complete frozen proposal
union; the exact attempt, proposal, job, and arm censuses are in the
online supplement. Conditional on an exactly recovered five-factor
structure in the clean directional populations, the hard-BIC contrast
identified the generating side with \(.987\)--\(.990\) accuracy across
the three profiles, whereas end-to-end yield was only
\(.403\)--\(.537\). Conditional Step 2 is therefore close to
deterministic but not perfect, while Step 1 limits the complete process.
Table~\ref{tbl-s3step2} gives the per-population incidence at the
default.

\begin{longtable}[]{@{}llrr@{}}
\caption{Step-2 preference at the default profile, by population:
bifactor-favoring incidence among available pairs and the median signed
contrast.}\label{tbl-s3step2}\tabularnewline
\toprule\noalign{}
Population & Truth role & Bifactor-favoring & Median \(\Delta\)BIC \\
\midrule\noalign{}
\endfirsthead
\toprule\noalign{}
Population & Truth role & Bifactor-favoring & Median \(\Delta\)BIC \\
\midrule\noalign{}
\endhead
\bottomrule\noalign{}
\endlastfoot
P1 & directional, bifactor & 598/598 = 1.000 & +1,088 \\
P1 + minor & robustness, bifactor major & 591/591 = 1.000 & +971 \\
P1 + doublet & robustness, bifactor major & 597/597 = 1.000 & +1,281 \\
P2 & directional, bifactor & 516/552 = .935 & +255 \\
P6 & directional, oblique & 20/595 = .034 & \(-72\) \\
P5 & parsimony preference & 10/575 = .017 & \(-53\) \\
P5 + doublet & LD hazard & 67/585 = .115 & \(-26\) \\
\end{longtable}

{\footnotesize\emph{Note.}
\(\Delta\mathrm{BIC} = \mathrm{BIC}_{\mathrm{oblique}} - \mathrm{BIC}_{\mathrm{bifactor}}\);
positive favors bifactor. Directional-accuracy language applies only to
P1, P2, and P6 under exact-five/full-recovery eligibility; the
robustness rows omit generated nuisance structure from both arms, and
the P5 row scores a parsimony preference among covariance-equivalent
representations. Full denominators are in the online supplement.}

Three readings follow (Table~\ref{tbl-s3step2}). First, median
directions and large majorities are strong in the expected
clean-direction cases, and the practical-tie band used for held-out
application reporting (\(\lvert\Delta\mathrm{BIC}\rvert \le 2\))
captures at most \(.022\) of available pairs in any population at the
default, with no exact tie. Second, the weak-core hazard corrupts Step-1
recovery rather than the verdict: despite P2's \(.455\) Step-1
precision, the great majority of its available pairs favored the
generating bifactor side (\(.945\) under exact-five eligibility). Third,
the absorbed-dependence hazard appears at the frozen rules on fresh data
in quantified form: omitted local dependence pulls \(.115\) of the
adversarial population's pairs to a bifactor-favoring contrast that is
not directional accuracy. At the overcomplete count
(\(K_{\mathrm{step2}} = 6\)), \(.593\) of the default profile's pairs
favor bifactor with a median contrast of \(+286\); these are
equivalence/absorption contrasts, reported as count-stratified
preference rather than accuracy.

\textbf{Guidance for the process.} Table~\ref{tbl-s3guide} states the
persistence and Step-2 rules frozen before confirmation, together with
the later descriptive count convention.

\begin{longtable}[]{@{}
  >{\raggedright\arraybackslash}p{(\linewidth - 4\tabcolsep) * \real{0.3333}}
  >{\raggedright\arraybackslash}p{(\linewidth - 4\tabcolsep) * \real{0.5333}}
  >{\raggedright\arraybackslash}p{(\linewidth - 4\tabcolsep) * \real{0.1333}}@{}}
\caption{Guidance for the process: persistence and Step 2 were frozen
before confirmation; the dual/full-window count convention is a post hoc
descriptive extension.}\label{tbl-s3guide}\tabularnewline
\toprule\noalign{}
\begin{minipage}[b]{\linewidth}\raggedright
Choice
\end{minipage} & \begin{minipage}[b]{\linewidth}\raggedright
Guidance
\end{minipage} & \begin{minipage}[b]{\linewidth}\raggedright
Basis
\end{minipage} \\
\midrule\noalign{}
\endfirsthead
\toprule\noalign{}
\begin{minipage}[b]{\linewidth}\raggedright
Choice
\end{minipage} & \begin{minipage}[b]{\linewidth}\raggedright
Guidance
\end{minipage} & \begin{minipage}[b]{\linewidth}\raggedright
Basis
\end{minipage} \\
\midrule\noalign{}
\endhead
\bottomrule\noalign{}
\endlastfoot
Backbone specification & anchor-only (AO) at both steps; AZ deliveries
are stable yet wrong at \(.44\)--\(.58\) on this grid & development
comparison above \\
Count evidence & ELBO and BIC gains at the 20\% cut with sustain one
over the complete fitted window; record one-versus-both support
descriptively & post hoc count convention; Section \ref{sec-route} \\
Persistence profiles & report all three frozen readings,
\texttt{.85/r1}, \texttt{.80/r2}, \texttt{.70/r3}; \texttt{.80/r2} is
the single practical default, retained under held-out confirmation &
development freeze; Section \ref{sec-s3-val} \\
Reading the three profiles & when all three deliver the same \(K_p\),
report it; when all deliver but \(K_p\) differs, report the delivered
range and claim a shared core only where one is directly verified; when
delivery status itself differs, report profile sensitivity (among
held-out joint-delivery count disagreements, \(.978\) are adjacent) &
Section \ref{sec-s3-val} \\
Step-2 comparison & two fixed arms at \(K_{\mathrm{step2}} = K_p\) under
the delivered design; hard BIC decides, with
\(\lvert\Delta\mathrm{BIC}\rvert \le 2\) a practical-tie band used for
held-out application reporting & Section \ref{sec-s3-step2} \\
Hazards & a weak core yields stable-yet-wrong counts that no tested
cutoff on the developed grid repairs (report per population); absorbed
local dependence pulls the step-2 contrast toward the general factor
while every step-1 indicator stays clean & P2; development and held-out
results; Section \ref{sec-s3-step2} \\
\end{longtable}

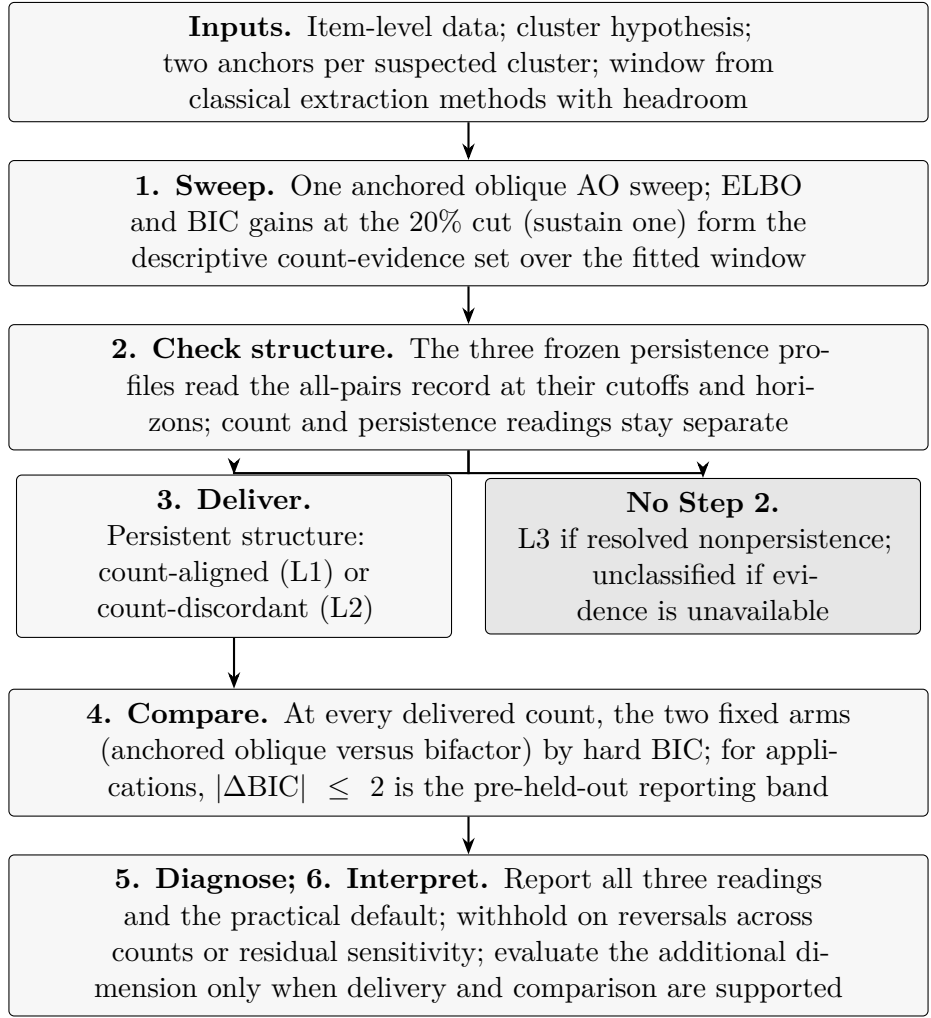
\begin{figure}[htbp]
\centering
\begin{tikzpicture}[
  node distance = 5mm and 8mm,
  box/.style  = {draw, rounded corners = 2pt, align = center, fill = gray!6,
                 inner sep = 5pt, text width = 118mm, font = \small},
  half/.style = {draw, rounded corners = 2pt, align = center,
                 inner sep = 5pt, text width = 54mm, font = \small},
  arr/.style  = {-{Stealth[length=2.2mm]}, thick}
]
\node[box] (inp) {\textbf{Inputs.} Item-level data; cluster hypothesis; two anchors per suspected cluster; window from classical extraction methods with headroom};
\node[box, below = of inp] (s1) {\textbf{1. Sweep.} One anchored oblique AO sweep; ELBO and BIC gains at the 20\% cut (sustain one) form the descriptive count-evidence set over the fitted window};
\node[box, below = of s1] (stab) {\textbf{2. Check structure.} The three frozen persistence profiles read the all-pairs record at their cutoffs and horizons; count and persistence readings stay separate};
\node[half, fill = gray!6]  (del)  at ($(stab.south)+(-31mm,-14mm)$) {\textbf{3. Deliver.}\\ Persistent structure: count-aligned (L1) or count-discordant (L2)};
\node[half, fill = gray!20] (hold) at ($(stab.south)+(31mm,-14mm)$) {\textbf{No Step 2.}\\ L3 if resolved nonpersistence;\\ unclassified if evidence is unavailable};
\node[box] (s2) at ($(stab.south)+(0,-40mm)$) {\textbf{4. Compare.} At every delivered count, the two fixed arms (anchored oblique versus bifactor) by hard BIC; for applications, $\lvert\Delta\mathrm{BIC}\rvert \le 2$ is the pre-held-out reporting band};
\node[box, below = of s2] (rep) {\textbf{5. Diagnose; 6. Interpret.} Report all three readings and the practical default; withhold on reversals across counts or residual sensitivity; evaluate the additional dimension only when delivery and comparison are supported};
\draw[arr] (inp) -- (s1);
\draw[arr] (s1) -- (stab);
\draw[arr] (stab.south) -- ++(0,-3mm) -| (del.north);
\draw[arr] (stab.south) -- ++(0,-3mm) -| (hold.north);
\draw[arr] (del.south) -- (del.south |- s2.north);
\draw[arr] (s2) -- (rep);
\end{tikzpicture}
\caption{The two-step rule and process as six steps: sweep, structure check, delivery branch, operational comparison, diagnosis, and interpretation. Identifiability permits estimation of each candidate, count selection proposes where to look, direct wider-count persistence determines whether a structure can be delivered, and the conditional comparison supplies evidence about distinguishability. Resolved all-source nonpersistence is L3 even if both count paths are unusable; unresolved persistence, or persistent $K_p$ when both paths are unusable, is unclassified.}
\label{fig:workflow}
\end{figure}

Figure \ref{fig:workflow} condenses the process into six actions--sweep,
check, deliver, compare, diagnose, and interpret--and
Table~\ref{tbl-s3guide} supplies their specifics. No single selection or
fit number stands alone.

\section{Empirical applications}\label{sec-empirical}

\subsection{Datasets, windows, and procedure}\label{sec-datasets}

PEFA RMSEA, SRMR, CFI, and TLI are established variational counterparts
of familiar SEM fit indices (\citeproc{ref-chenjin2026fit}{Chen \& Jin,
2026a}). These ancillary conservative screens are weakly discriminating
here: 18 of the 25 fitted candidates, including all four dual-criterion
count readings, pass RMSEA \(\le .06\), SRMR \(\le .10\), CFI
\(\ge .90\), and TLI \(\ge .90\) (Table S18). They inform interpretation
rather than determine delivery; count and structural delivery follow the
gain and persistence rules.

Four empirical datasets span the evidence spectrum
(Table~\ref{tbl-datasets}). All items are treated as continuous, with
complete cases and standard reverse-keying for the bfi. The nine-test
Holzinger--Swineford excerpt common in software examples is avoided
deliberately. With \(K = 3\), the higher-order structure is a saturated
reparameterization of \(\Phi\) (three correlations, three \(\gamma_k\)),
so the HO-versus-oblique distinction is untestable. The full battery
restores degrees of freedom.

\begin{longtable}[]{@{}
  >{\raggedright\arraybackslash}p{(\linewidth - 10\tabcolsep) * \real{0.1304}}
  >{\raggedright\arraybackslash}p{(\linewidth - 10\tabcolsep) * \real{0.1304}}
  >{\raggedleft\arraybackslash}p{(\linewidth - 10\tabcolsep) * \real{0.1739}}
  >{\centering\arraybackslash}p{(\linewidth - 10\tabcolsep) * \real{0.2174}}
  >{\centering\arraybackslash}p{(\linewidth - 10\tabcolsep) * \real{0.2174}}
  >{\raggedright\arraybackslash}p{(\linewidth - 10\tabcolsep) * \real{0.1304}}@{}}
\caption{The four empirical
datasets.}\label{tbl-datasets}\tabularnewline
\toprule\noalign{}
\begin{minipage}[b]{\linewidth}\raggedright
Dataset
\end{minipage} & \begin{minipage}[b]{\linewidth}\raggedright
Source
\end{minipage} & \begin{minipage}[b]{\linewidth}\raggedleft
\(N\)
\end{minipage} & \begin{minipage}[b]{\linewidth}\centering
\(J \times K\)
\end{minipage} & \begin{minipage}[b]{\linewidth}\centering
Window (\(K_0\); sources)
\end{minipage} & \begin{minipage}[b]{\linewidth}\raggedright
Description
\end{minipage} \\
\midrule\noalign{}
\endfirsthead
\toprule\noalign{}
\begin{minipage}[b]{\linewidth}\raggedright
Dataset
\end{minipage} & \begin{minipage}[b]{\linewidth}\raggedright
Source
\end{minipage} & \begin{minipage}[b]{\linewidth}\raggedleft
\(N\)
\end{minipage} & \begin{minipage}[b]{\linewidth}\centering
\(J \times K\)
\end{minipage} & \begin{minipage}[b]{\linewidth}\centering
Window (\(K_0\); sources)
\end{minipage} & \begin{minipage}[b]{\linewidth}\raggedright
Description
\end{minipage} \\
\midrule\noalign{}
\endhead
\bottomrule\noalign{}
\endlastfoot
ICAR ability & Condon \& Revelle (\citeproc{ref-condon2014icar}{2014});
psychTools & 1,248 & \(16 \times 4\) & \(2{:}7\) (\(2\); \(2{:}4\)) &
cognitive ability items, four content domains \\
Holzinger--Swineford, 24 tests & Holzinger \& Swineford
(\citeproc{ref-holzinger1939study}{1939}); psychTools & 301 &
\(24 \times 5\) & \(3{:}8\) (\(3\); \(3{:}5\)) & classic multi-domain
test battery \\
PID-5 domain-primary facets & Krueger et al.
(\citeproc{ref-krueger2012pid5}{2012}); Roskam et al.
(\citeproc{ref-roskam2015pid5}{2015}) & 2,532 & \(15 \times 5\) &
\(3{:}8\) (\(3\); \(3{:}5\)) & maladaptive personality facets, five
domains \\
bfi & psychTools (\citeproc{ref-revelle2024psych}{Revelle, 2024}) &
2,436 & \(25 \times 5\) & \(3{:}9\) (\(3\); \(3{:}6\)) & five-domain
personality items, contested structure \\
\end{longtable}

{\footnotesize\emph{Note.} \(K\) counts theoretical clusters; the counts
the process selects and delivers are reported in Sections
\ref{sec-sweeps} and \ref{sec-transition} and need not match. Complete
cases; all items treated as continuous in the primary analyses (the ICAR
items are dichotomous and the bfi items ordinal, so a categorical
estimator is a feasibility question the Discussion returns to); bfi
reverse-keyed on its standard seven items. PID-5 facet scores are means
of four items over the fifteen domain-primary facets (three per DSM-5
domain).}

The applications use the same fitted sweeps and persistence profiles as
the frozen process of Section \ref{sec-study3}, and the analyses were
completed before the held-out simulation stage opened, so they are
independent of its outcome. Each dataset uses the anchor-only backbone
with the first two instrument-order items per prespecified cluster as
anchors, one oblique sweep over its declared window
(Table~\ref{tbl-datasets}), and the three frozen reference readings. The
publication rereading applies the ELBO and BIC 20\%/sustain-one paths to
each complete fitted window. Each window brackets the classical
factor-number range reported in the online supplement with headroom
above. Two regularities in those classical results are noteworthy.
First, eigenvalue-based methods return fewer factors than the
theoretical cluster count when a strong general factor flattens the
tail. Second, on the Holzinger battery all methods agree on 4, not the
classic 5. The reporting hierarchy follows the decision path: the sweep
table and the all-pairs persistence matrices are the primary evidence,
since every delivery decision reads only those two, while the
adjacent-transition diagnostics are auxiliary. For empirical data,
truth-based accuracy does not exist: the readings below are
count-conditional descriptions.

\subsection{Step 1: the sweep and count evidence}\label{sec-sweeps}

Table~\ref{tbl-emptrans} compacts the step-1 record into the gains and
adjacent stability indices at every step; the full sweep and diagnostic
record is in the online supplement.

\begin{longtable}[]{@{}
  >{\raggedright\arraybackslash}p{(\linewidth - 12\tabcolsep) * \real{0.2182}}
  >{\raggedright\arraybackslash}p{(\linewidth - 12\tabcolsep) * \real{0.0909}}
  >{\raggedleft\arraybackslash}p{(\linewidth - 12\tabcolsep) * \real{0.1455}}
  >{\raggedleft\arraybackslash}p{(\linewidth - 12\tabcolsep) * \real{0.1273}}
  >{\raggedleft\arraybackslash}p{(\linewidth - 12\tabcolsep) * \real{0.1091}}
  >{\raggedleft\arraybackslash}p{(\linewidth - 12\tabcolsep) * \real{0.1636}}
  >{\centering\arraybackslash}p{(\linewidth - 12\tabcolsep) * \real{0.1455}}@{}}
\caption{Step-1 transitions under the frozen anchor-only frame:
scale-free gains and adjacent stability indices at every
step.}\label{tbl-emptrans}\tabularnewline
\toprule\noalign{}
\begin{minipage}[b]{\linewidth}\raggedright
Dataset
\end{minipage} & \begin{minipage}[b]{\linewidth}\raggedright
Step
\end{minipage} & \begin{minipage}[b]{\linewidth}\raggedleft
ELBO g\%
\end{minipage} & \begin{minipage}[b]{\linewidth}\raggedleft
BIC g\%
\end{minipage} & \begin{minipage}[b]{\linewidth}\raggedleft
\(\phi_{\min}\)
\end{minipage} & \begin{minipage}[b]{\linewidth}\raggedleft
RMSD max
\end{minipage} & \begin{minipage}[b]{\linewidth}\centering
Collision
\end{minipage} \\
\midrule\noalign{}
\endfirsthead
\toprule\noalign{}
\begin{minipage}[b]{\linewidth}\raggedright
Dataset
\end{minipage} & \begin{minipage}[b]{\linewidth}\raggedright
Step
\end{minipage} & \begin{minipage}[b]{\linewidth}\raggedleft
ELBO g\%
\end{minipage} & \begin{minipage}[b]{\linewidth}\raggedleft
BIC g\%
\end{minipage} & \begin{minipage}[b]{\linewidth}\raggedleft
\(\phi_{\min}\)
\end{minipage} & \begin{minipage}[b]{\linewidth}\raggedleft
RMSD max
\end{minipage} & \begin{minipage}[b]{\linewidth}\centering
Collision
\end{minipage} \\
\midrule\noalign{}
\endhead
\bottomrule\noalign{}
\endlastfoot
ability & 2--3 & 100.0 & 100.0 & .042 & .430 & No \\
ability & 3--4 & -2.3 & -18.2 & .880 & .167 & No \\
ability & 4--5 & -137.4 & -5.6 & .940 & .073 & No \\
ability & 5--6 & -220.2 & 11.3 & 1.000 & .005 & No \\
ability & 6--7 & -333.7 & -206.5 & .753 & .178 & No \\
Holzinger 24 & 3--4 & 100.0 & 100.0 & .879 & .180 & No \\
Holzinger 24 & 4--5 & -68.2 & -33.8 & .964 & .092 & No \\
Holzinger 24 & 5--6 & -45.1 & 10.7 & .885 & .115 & No \\
Holzinger 24 & 6--7 & -95.5 & -4.4 & .859 & .103 & No \\
Holzinger 24 & 7--8 & -168.6 & -320.7 & .216 & .256 & Yes \\
PID-5 & 3--4 & 100.0 & 100.0 & .937 & .148 & No \\
PID-5 & 4--5 & 47.8 & 41.2 & .951 & .109 & No \\
PID-5 & 5--6 & 2.5 & 10.7 & .896 & .126 & No \\
PID-5 & 6--7 & -4.7 & 0.2 & .773 & .168 & No \\
PID-5 & 7--8 & -11.0 & -24.2 & .697 & .143 & No \\
bfi & 3--4 & 86.3 & 81.9 & .347 & .365 & No \\
bfi & 4--5 & 100.0 & 100.0 & .949 & .106 & No \\
bfi & 5--6 & 29.6 & 28.6 & .767 & .209 & No \\
bfi & 6--7 & 10.0 & 9.4 & .803 & .196 & No \\
bfi & 7--8 & 1.2 & 3.4 & .618 & .216 & Yes \\
bfi & 8--9 & -3.7 & -21.6 & .686 & .192 & No \\
\end{longtable}

{\footnotesize\emph{Note.} Steps are labelled by the counts they join.
Gains are percentages of the largest single-step gain in that window.
\(\phi_{\min}\) is minimum absolute Tucker congruence over the
minimum-SSE matched columns; RMSD max is the worst RMSD over the same
matches. Pairwise sign alignment makes both reflection-invariant.
Collision = Yes means two source columns independently selected the same
target; that edge remains descriptive but is vetoed for persistence.}

Both count paths resolve the same singleton set in every application:
\(C_{20}(W)=\{3\}\) for ability, \(\{4\}\) for Holzinger 24, \(\{5\}\)
for the PID-5, and \(\{6\}\) for the bfi. These are descriptive count
readings; delivery belongs to the persistence profiles of the next
subsection.

\subsection{The persistence profiles and their
deliveries}\label{sec-transition}

An adjacent step is a local check. Table~\ref{tbl-emppers} widens it to
every larger candidate in the window, and the three frozen profiles read
it at their declared cutoffs and horizons.

\begin{longtable}[]{@{}lrrrrrrr@{}}
\caption{Direct all-pairs minimum absolute congruence from each source
structure to every larger target candidate (ability, Holzinger 24, and
PID-5; the bfi matrix is in the online
supplement).}\label{tbl-emppers}\tabularnewline
\toprule\noalign{}
Dataset & Source \(K\) & 3 & 4 & 5 & 6 & 7 & 8 \\
\midrule\noalign{}
\endfirsthead
\toprule\noalign{}
Dataset & Source \(K\) & 3 & 4 & 5 & 6 & 7 & 8 \\
\midrule\noalign{}
\endhead
\bottomrule\noalign{}
\endlastfoot
ability & 2 & .042 & .041 & .040 & .040 & .039 & \\
ability & 3 & & .880 & .878 & .879 & .886 & \\
ability & 4 & & & .940 & .940 & .965 & \\
ability & 5 & & & & 1.000 & .763 & \\
ability & 6 & & & & & .753 & \\
Holzinger 24 & 3 & & .879 & .874 & .797 & .758 & .666 \\
Holzinger 24 & 4 & & & .964 & .869 & .837 & .744 \\
Holzinger 24 & 5 & & & & .885 & .873 & .725 \\
Holzinger 24 & 6 & & & & & .859 & .162 \\
Holzinger 24 & 7 & & & & & & .216 \\
PID-5 & 3 & & .937 & .844 & .850 & .789 & .813 \\
PID-5 & 4 & & & .951 & .918 & .884 & .886 \\
PID-5 & 5 & & & & .896 & .791 & .819 \\
PID-5 & 6 & & & & & .773 & .792 \\
PID-5 & 7 & & & & & & .697 \\
\end{longtable}

{\footnotesize\emph{Note.} Rows identify the source structure and
columns the larger target solution. Ability targets span 3--7; Holzinger
24 and PID-5 targets span 4--8. Entries for \(K \rightarrow K+1\)
reproduce the corresponding adjacent step of Table~\ref{tbl-emptrans};
blanks fall outside the dataset's window or are not wider-count
comparisons.}

Table~\ref{tbl-empreadings} gives the resulting readings. For
\textbf{ability} all three profiles deliver at four, one step above the
count reading, because both the four-factor structure and the
three-factor core persist against every larger candidate, so the rule
delivers the larger persistent structure and carries the count
uncertainty forward. For \textbf{Holzinger 24} all three deliver at
five, above a count reading of four, before the over-extracted region
collapses. For the \textbf{PID-5} all three profiles deliver: the
shallow and deep readings give five, whereas the default gives four
because the \(5\rightarrow7\) congruence is \(.791\), only \(.009\)
below its \(.80\) cutoff. The four-factor core is therefore robust
across profiles, while the fifth dimension is marginal and
threshold-sensitive, not a third formal \(K_p=5\) reading; both
delivered counts enter Step 2. For the \textbf{bfi} only the shallow
profile delivers, at four and two counts below \(C_{20}(W)=\{6\}\),
while the default and deep profiles return no stable structure. Its gap
of two sits at the outer edge of the empirical record: it locates
disagreement but does not corroborate delivery. Persistence delivers two
counts below \(C_{20}(W)\); because empirical truth is unknown, this is
a below-count caution. In held-out truth-labelled data, the matching
below/gap-two cell recovers the generating structure in none of its 23
cases, reinforcing that caution.

The frozen source set is common to all three profiles and sized for the
deepest, \(r=3\) reading, so the shallower profiles deliberately do not
search a larger source universe. Eight of the twelve readings land on
that ceiling. Each application was therefore re-read from the widest
profile-specific source set its fitted window permits, without refitting
(online supplement, Table S22). The default is unchanged in all four
batteries, while the deep profile cannot be extended in any of them, its
frozen set already being the widest its window allows at \(r=3\). Only
the shallow reading moves: ability rises from four to five and Holzinger
24 from five to six. Thus a ceiling reading can be, but need not be,
truncated. The frozen readings remain primary; this post hoc sensitivity
shows that shallow persistence is most dependent on the declared source
ceiling.

\begin{longtable}[]{@{}
  >{\raggedright\arraybackslash}p{(\linewidth - 10\tabcolsep) * \real{0.1500}}
  >{\raggedleft\arraybackslash}p{(\linewidth - 10\tabcolsep) * \real{0.2000}}
  >{\raggedleft\arraybackslash}p{(\linewidth - 10\tabcolsep) * \real{0.2000}}
  >{\raggedright\arraybackslash}p{(\linewidth - 10\tabcolsep) * \real{0.1500}}
  >{\raggedright\arraybackslash}p{(\linewidth - 10\tabcolsep) * \real{0.1500}}
  >{\raggedright\arraybackslash}p{(\linewidth - 10\tabcolsep) * \real{0.1500}}@{}}
\caption{The three frozen reference readings on the four datasets: the
dual-criterion count-evidence set, persistence-selected count, and
revised layer.}\label{tbl-empreadings}\tabularnewline
\toprule\noalign{}
\begin{minipage}[b]{\linewidth}\raggedright
Dataset
\end{minipage} & \begin{minipage}[b]{\linewidth}\raggedleft
\(n\)
\end{minipage} & \begin{minipage}[b]{\linewidth}\raggedleft
\(C_{20}(W)\)
\end{minipage} & \begin{minipage}[b]{\linewidth}\raggedright
\texttt{.85/r1}
\end{minipage} & \begin{minipage}[b]{\linewidth}\raggedright
\texttt{.80/r2}
\end{minipage} & \begin{minipage}[b]{\linewidth}\raggedright
\texttt{.70/r3}
\end{minipage} \\
\midrule\noalign{}
\endfirsthead
\toprule\noalign{}
\begin{minipage}[b]{\linewidth}\raggedright
Dataset
\end{minipage} & \begin{minipage}[b]{\linewidth}\raggedleft
\(n\)
\end{minipage} & \begin{minipage}[b]{\linewidth}\raggedleft
\(C_{20}(W)\)
\end{minipage} & \begin{minipage}[b]{\linewidth}\raggedright
\texttt{.85/r1}
\end{minipage} & \begin{minipage}[b]{\linewidth}\raggedright
\texttt{.80/r2}
\end{minipage} & \begin{minipage}[b]{\linewidth}\raggedright
\texttt{.70/r3}
\end{minipage} \\
\midrule\noalign{}
\endhead
\bottomrule\noalign{}
\endlastfoot
ability & 1,248 & 3 & \(K_p=4^{\dagger}\), L2 & \(K_p=4\), L2 &
\(K_p=4\), L2 \\
Holzinger 24 & 301 & 4 & \(K_p=5^{\dagger}\), L2 & \(K_p=5\), L2 &
\(K_p=5\), L2 \\
PID-5 & 2,532 & 5 & \(K_p=5\), L1 & \(K_p=4\), L2 & \(K_p=5\), L1 \\
bfi & 2,436 & 6 & \(K_p=4\), L2 & none, L3 & none, L3 \\
\end{longtable}

{\footnotesize\emph{Note.} ELBO and BIC give the displayed singleton in
all four datasets. Consequently every empirical L1 cell has support from
both criteria and every L2 cell from neither; among L2 readings,
\(g_{\mathrm{CP}}=1\) except for the bfi shallow L2 reading, where it is
2. PID-5 cannot be assigned a third \(K_p=5\) L2 reading: if five passed
the default profile it would belong to L1 because \(C_{20}(W)=\{5\}\).
\(\dagger\) The widest profile-specific sensitivity reading is
respectively five and six; the frozen common-source reading remains
primary.}

\subsection{Step 2: the comparison at the delivered
structures}\label{sec-step2}

Table~\ref{tbl-empstep2} poses the frozen two-arm comparison at every
distinct delivered count, with \(K_{\mathrm{step2}} = K_p\) and both
arms sharing the inherited design of Section \ref{sec-route}.

\begin{longtable}[]{@{}
  >{\raggedright\arraybackslash}p{(\linewidth - 8\tabcolsep) * \real{0.1667}}
  >{\raggedleft\arraybackslash}p{(\linewidth - 8\tabcolsep) * \real{0.2222}}
  >{\raggedleft\arraybackslash}p{(\linewidth - 8\tabcolsep) * \real{0.2222}}
  >{\raggedleft\arraybackslash}p{(\linewidth - 8\tabcolsep) * \real{0.2222}}
  >{\raggedright\arraybackslash}p{(\linewidth - 8\tabcolsep) * \real{0.1667}}@{}}
\caption{Step-2 anchored comparisons at the delivered structures: hard
BIC for the two fixed arms and the signed
contrast.}\label{tbl-empstep2}\tabularnewline
\toprule\noalign{}
\begin{minipage}[b]{\linewidth}\raggedright
Dataset (\(K_{\mathrm{step2}}\))
\end{minipage} & \begin{minipage}[b]{\linewidth}\raggedleft
BIC oblique
\end{minipage} & \begin{minipage}[b]{\linewidth}\raggedleft
BIC bifactor
\end{minipage} & \begin{minipage}[b]{\linewidth}\raggedleft
\(\Delta\)BIC
\end{minipage} & \begin{minipage}[b]{\linewidth}\raggedright
Preference
\end{minipage} \\
\midrule\noalign{}
\endfirsthead
\toprule\noalign{}
\begin{minipage}[b]{\linewidth}\raggedright
Dataset (\(K_{\mathrm{step2}}\))
\end{minipage} & \begin{minipage}[b]{\linewidth}\raggedleft
BIC oblique
\end{minipage} & \begin{minipage}[b]{\linewidth}\raggedleft
BIC bifactor
\end{minipage} & \begin{minipage}[b]{\linewidth}\raggedleft
\(\Delta\)BIC
\end{minipage} & \begin{minipage}[b]{\linewidth}\raggedright
Preference
\end{minipage} \\
\midrule\noalign{}
\endhead
\bottomrule\noalign{}
\endlastfoot
ability (4) & 53,068.7 & 53,019.5 & +49.2 & bifactor \\
Holzinger 24 (5) & 18,237.9 & 18,236.1 & +1.8 & practical tie \\
PID-5 (4) & 96,716.4 & 96,277.8 & +438.6 & bifactor \\
PID-5 (5) & 96,289.8 & 96,308.8 & -19.0 & oblique \\
bfi (4) & 158,581.8 & 157,108.6 & +1,473.2 & bifactor \\
\end{longtable}

{\footnotesize\emph{Note.}
\(\Delta\mathrm{BIC} = \mathrm{BIC}_{\mathrm{oblique}} - \mathrm{BIC}_{\mathrm{bifactor}}\);
positive favors bifactor; \(\lvert\Delta\mathrm{BIC}\rvert \le 2\) is
the pre-held-out amendment's application-reporting band. Every delivered
count from Table~\ref{tbl-empreadings} is fitted once; the bfi
comparison exists only because its shallow reading delivered.}

The four count-conditional patterns offer qualitative analogies, not
replications of truth-labelled mechanisms (Table~\ref{tbl-empstep2}).
Ability combines an above-path delivery with bifactor support. Holzinger
24 is a practical tie at its delivered count; that tie does not itself
support a higher-order or reducibility interpretation. Its fitted
bifactor arm is anatomically mixed rather than the proportional
simple-structure form of Proposition 1: ECV is \(.523\), only one
cluster approaches proportionality, and two hypothesized group factors
collapse (online supplement, Tables S23 and S24). The PID-5 changes sign
across its two delivered counts, reinforcing the count-sensitive
boundary reading. For the bfi, the shallow L2 delivery--not its
gap--properly opens Step 2 at \(K=4\), where BIC strongly favors
bifactor (\(\Delta\mathrm{BIC}=+1{,}473.2\)). Both deeper profiles
reject delivery, so this is a shallow/below-count caution rather than
robust bifactor evidence. These are four distinct descriptive
cases--support, a practical tie, count-conditional reversal, and support
that survives only at a count the two deeper profiles refused--not four
truth classifications. The delivered \(K=4\) solutions for PID-5 and bfi
also fail the ancillary fit screens (Table S18), so their Step-2
contrasts carry particular caution about absolute fit.

\section{Discussion}\label{sec-discussion}

Selection is not delivery: gain readings identify a count neighborhood,
while persistence determines whether a structure is interpretable.
Preliminary work supplies tentative components, development selects the
profile family, held-out confirmation estimates same-generator
performance, and empirical cases only illustrate it. Although the frozen
ELBO path selects the generating count in 97.4\% of held-out datasets,
the weak-core population still yields stable but wrong deliveries.
Across the shared six-population domain, all three profiles deliver in
3,471/3,600 datasets (96.4\%) but agree on exact \(K_p\) in only
2,399/3,471 (69.1\%): delivery status is robust across the profile
family, whereas the boundary count is sensitive, and neither establishes
truth. The theory supplies a compatible population-level frame, subject
to Theorem 1's conditions: distinguishability disappears at the
reducible boundary (Proposition 1), holds when every cluster is
non-proportional (Theorem 1), and remains pattern-dependent in mixed
configurations.

We extend the PEFA-fit gain approach (\citeproc{ref-chenjin2026fit}{Chen
\& Jin, 2026a}) by treating either ELBO or BIC gain at the 20\% cut with
sustain one as descriptive count support over the complete fitted
window. Here \(C_{20}(W)\) describes that neighborhood, while
persistence determines delivery and \(K_p\); one-versus-both support and
a positive L2 gap only record provenance and locality. Because the
dual/full-window partition was post hoc, unlike the fits and persistence
profiles, it requires independent confirmation.

The analytical and persistence layers are not interchangeable. \(D_K\)
is a population covariance distance; \(\phi_{\min}\) and
\(\mathrm{RMSD}_{\max}\) compare fitted loading columns, with a cutoff
controlling tolerated reorganization and \(r\) the number of wider
checks. Neither estimates \(D_K\). Persistence supplies the structure on
which Step 2 asks necessity; it does not answer that question, and
profile disagreement is a local stability warning rather than a distance
scale.

Two hazards share one mechanism: covariance omitted from the first-order
structure can flow into the general column. Preliminary Finding 6
demonstrates this after merging a factor; the adversarial doublet does
so at the correct count, with held-out bifactor-favoring rates of
\(.045\)--\(.154\) despite clean stability indicators. The latter is
harder because count stability cannot detect residual dependence. This
designed doublet establishes possibility, not an empirical diagnosis,
and data-selected residual pairs add post-selection uncertainty. Where
diagonal uniqueness is doubtful, report residual diagnostics and develop
calibrated sparse-residual extensions (\citeproc{ref-jin2026sparse}{Jin
et al., 2026}).

Neighboring work spans bi-factor rotation
(\citeproc{ref-jennrich2011exploratory}{Jennrich \& Bentler, 2011},
\citeproc{ref-jennrich2012exploratory}{2012}), empirical targets
(\citeproc{ref-garciagarzon2019improving}{Garcia-Garzon et al., 2019};
\citeproc{ref-jimenez2023exploratory}{Jiménez et al., 2023}), recovery
guarantees (\citeproc{ref-qiao2025exact}{Qiao et al., 2025a},
\citeproc{ref-qiao2025hierarchical}{2025b}), and fixed-pattern
identifiability (\citeproc{ref-fang2021identifiability}{Fang et al.,
2021}; \citeproc{ref-peeters2012rotational}{Peeters, 2012}), but does
not cross anchoring with reducibility. Existing guarantees assume
stronger blockwise non-proportionality, excluding reducible and
silent-cluster cases where partial specification matters. The
second-order/bi-factor EFA unification
(\citeproc{ref-asparouhov2026unification}{Asparouhov \& Muthén, 2026})
likewise treats the structures as near twins, but describes a
representation at a supplied count; the anchored workflow instead
selects and then compares structures.

Six choices summarize practice. \emph{First}, within the studied
envelope use an oblique rather than bifactor sweep: mild oblique
over-extraction retains the core, whereas bifactor under-counting is
more damaging. \emph{Second}, use anchor-only at both steps; anchor-zero
often gives stable but wrong bifactor-population deliveries.
\emph{Third}, read count and persistence separately, report all three
profiles, use \texttt{.80/r2} as the practical default, and treat
disagreement as boundary uncertainty. \emph{Fourth}, report recovery by
population because persistence need not be correct. \emph{Fifth}, after
L1/L2 delivery compare the two Step-2 arms at
\(K_{\mathrm{step2}}=K_p\); this is conditional evidence, not an
estimate of \(D_K\). \emph{Sixth}, report resolved nonpersistence as
non-delivery and unresolved evidence as unclassified; never fall back to
an undelivered count, and inspect local dependence before interpreting a
general factor.

A frozen-structure maximum-likelihood refit is not a selection check: it
reweights the hard-selected pattern but cannot recover omitted loadings.
Quantifying how it inherits this omission-dominated error profile
requires generating truth and belongs to a separate study.

The simulations use fixed four-item clusters, correct anchors, a
stylized doublet, and five-major-factor targets. Thus L3 measures
non-delivery relative to that target, not sensitivity to a population
with no stable core. Held-out results reproduce the same generator
families, not external populations; they do not establish universal
cutoffs, and surplus diagnostics remain descriptive. Step 2 is an
anchored variational-BIC comparison conditional on delivery, not a test
of \(D_K\). Future work should vary anchors, cluster hypotheses and
sizes, include true no-core populations, calibrate surplus and residual
diagnostics, develop post-selection and boundary-aware inference, extend
categorical estimation (\citeproc{ref-chen2020pcirt}{Chen, 2020}), and
study sparse residuals (\citeproc{ref-jin2026sparse}{Jin et al., 2026})
and persistence depth.

\section*{Data and Code Availability}\label{data-and-code-availability}
\addcontentsline{toc}{section}{Data and Code Availability}

A checksum-verified replication archive is available through OSF:
\url{https://osf.io/8y69s/overview?view_only=e90a4e18f936435aa45a14ce5c86ea5f}.
It contains scientific scripts, review-safe aggregate results,
table-reproduction code, and provenance records; its README documents
exclusions and the executed source fingerprint. The public datasets can
be reconstructed from psychTools. Only aggregate PID-5 results are
distributed because its responses and item wording cannot be
redistributed. The project will become public and citable upon
acceptance; \textbf{vbpm} 0.9.1 is available from CRAN, with tagged
source on GitHub.

\section*{References}\label{references}
\addcontentsline{toc}{section}{References}

\protect\phantomsection\label{refs}

\section*{Appendix A: Proof of Theorem
1}\label{appendix-a-proof-of-theorem-1}
\addcontentsline{toc}{section}{Appendix A: Proof of Theorem 1}

Throughout, \(\Sigma = C + \Psi\) with
\(C = \mathbf{b}_g\mathbf{b}_g' + \sum_k \mathbf{b}_s^k \mathbf{b}_s^{k\prime}\),
\(K \ge 2\) disjoint clusters, general loadings nonzero on every item,
\(n_k \ge 3\) items in cluster \(k\), and at least two nonzero group
loadings per cluster. Write \(\mathbf{v}_k = \mathbf{b}_{g|k}\) for the
general column's restriction to cluster \(k\), and recall the
decomposition
\(\mathbf{b}_{g|k} = \alpha_k \mathbf{u}_k + \mathbf{w}_k\) of Section
\ref{sec-disting}: cluster \(k\) is proportional exactly when
\(\mathbf{w}_k = \mathbf{0}\).

The argument has three moves. First, any rival \(K\)-factor
representation of \(\Sigma\) can differ from \(C\) only by a diagonal
matrix (Lemma A1). Second, rank is counted cluster by cluster: the rival
splits into a rank-one general term plus one block per cluster, and no
diagonal perturbation can flatten a non-proportional cluster's block
(Lemmas A2 and A4), so the \(K\) clusters exhaust a rank-\(K\) budget
exactly. Third, exact exhaustion forces every block's range onto the
restricted general column, which is precisely the proportional
configuration the hypothesis excludes (Proof of Theorem 1). The
workhorse is Lemma A3, an elementary fact about rank-one differences.

\textbf{Lemma A1 (diagonal difference).} Suppose \(\Sigma\) admits a
\(K\)-factor representation: \(\Sigma = \tilde{C} + \tilde{\Psi}\) with
\(\tilde{C}\) PSD of rank at most \(K\) and \(\tilde{\Psi}\) diagonal.
Then \(\tilde{C} = C + \Delta\) with \(\Delta = \Psi - \tilde{\Psi}\)
diagonal. So it suffices to show
\(\operatorname{rank}(C + \Delta) \ge K + 1\) for every diagonal
\(\Delta\).

\textbf{Lemma A2 (block form and rank additivity).} For any diagonal
\(\Delta\), \(C + \Delta = \mathbf{b}_g\mathbf{b}_g' + M\), where \(M\)
is block diagonal over clusters with blocks
\(M_k = \mathbf{b}_s^k \mathbf{b}_s^{k\prime} + \Delta_k\). Let
\(G_k^\perp = \{x \in \mathbb{R}^{n_k}: x \perp \mathbf{v}_k\}\). For
\(x \in G_k^\perp\) (embedded in the cluster-\(k\) coordinates),
\((C+\Delta)x = \mathbf{b}_g(\mathbf{v}_k'x) + M_k x = M_k x\), and
\(M_k x\) lies in the cluster-\(k\) coordinate block. Hence the column
space of \(C + \Delta\) contains \(\bigoplus_k M_k(G_k^\perp)\), whose
summands occupy disjoint coordinate blocks, so with
\(r_k = \dim M_k(G_k^\perp)\),
\(\operatorname{rank}(C + \Delta) \ge \sum_k r_k\).

\textbf{Lemma A3 (proportionality forcing).} Let
\(\mathbf{b}, \mathbf{s} \in \mathbb{R}^n\) with \(n \ge 3\) and
\(\mathbf{b}\) entrywise nonzero, and suppose
\(\tau\,\mathbf{b}\mathbf{b}' - \mathbf{s}\mathbf{s}'\) is diagonal for
some scalar \(\tau\). Then either \(\mathbf{s}\) has at most one nonzero
entry, or \(\mathbf{s} = \kappa\,\mathbf{b}\) with \(\kappa \ne 0\).
\emph{Proof.} The off-diagonal entries give \(\tau\, b_i b_j = s_i s_j\)
for all \(i \ne j\). If \(\tau = 0\), the pairwise products vanish, so
at most one \(s_i\) is nonzero. If \(\tau \ne 0\), no ratio
\(\rho_i = s_i/b_i\) is zero (every pairwise product equals \(\tau\)),
and \(\rho_i \rho_j = \tau\) for all pairs; fixing \(i\) and taking
\(j \ne l\) both distinct from \(i\) gives
\(\rho_i(\rho_j - \rho_l) = 0\), so all ratios other than \(\rho_i\)
share one value, and \(\rho_i \rho_j = \tau = \rho_j \rho_l\) with
\(\rho_j \ne 0\) folds \(\rho_i\) into the same value: all ratios are
equal and nonzero. \(\square\)

Applied within cluster \(k\) (with \(\mathbf{b} = \mathbf{v}_k\) and
\(\mathbf{s} = \mathbf{b}_s^k\)), the second alternative is exactly
\(\mathbf{w}_k = \mathbf{0}\), and the first is the single-indicator
spike; both are excluded for every cluster under Theorem 1's hypothesis.

\textbf{Lemma A4 (silence characterization).} For every diagonal
\(\Delta_k\), \(r_k = 0\) implies that cluster \(k\) is proportional
(\(\mathbf{w}_k = \mathbf{0}\)) or its group loading is a
single-indicator spike; conversely, each of those configurations admits
a diagonal \(\Delta_k\) with \(r_k = 0\). \emph{Proof.} \(r_k = 0\)
means \(M_k\) annihilates \(G_k^\perp\); a symmetric matrix annihilating
the orthocomplement of \(\mathbf{v}_k\) is a multiple
\(\tau\,\mathbf{v}_k\mathbf{v}_k'\), so
\(\Delta_k = \tau\,\mathbf{v}_k\mathbf{v}_k' - \mathbf{b}_s^k \mathbf{b}_s^{k\prime}\)
must be diagonal, and Lemma A3 gives the two alternatives. Conversely,
\(\Delta_k = \tau\,\mathbf{v}_k\mathbf{v}_k' - \mathbf{b}_s^k\mathbf{b}_s^{k\prime}\)
is itself diagonal in both configurations (with \(\tau = \kappa^2\) in
the proportional case and \(\tau = 0\) in the spike case), so both can
produce \(r_k = 0\). \(\square\)

\textbf{Proof of Theorem 1.} Let \(\Delta\) be diagonal and suppose
\(\operatorname{rank}(C + \Delta) \le K\); a contradiction follows in
three steps.

\emph{Step 1 (the rank budget is exactly spent).} Every cluster has
\(\mathbf{w}_k \ne \mathbf{0}\) and no spike, so \(r_k \ge 1\) for all
\(k\) by Lemma A4, while Lemma A2 gives \(\sum_k r_k \le K\). Hence
\(r_k = 1\) for every \(k\), and the column space of \(C+\Delta\) is
\emph{equal} to \(V = \bigoplus_k M_k(G_k^\perp)\), whose cluster-\(l\)
coordinate block is the one-dimensional \(M_l(G_l^\perp)\).

\emph{Step 2 (cross-cluster alignment).} Take any item \(j\) in any
cluster \(k\); its column of \(C + \Delta\) is
\(g_j \mathbf{b}_g + M_k e_j\) and must lie in \(V\). For any
\(l \ne k\), the column's cluster-\(l\) block is \(g_j \mathbf{v}_l\)
(the term \(M_k e_j\) is supported on cluster \(k\)), so membership in
\(V\) requires \(g_j \mathbf{v}_l \in M_l(G_l^\perp)\); since
\(g_j \ne 0\) and \(\mathbf{v}_l \ne \mathbf{0}\), the one-dimensional
space \(M_l(G_l^\perp)\) must equal
\(\operatorname{span}\{\mathbf{v}_l\}\). With \(K \ge 2\), such a column
exists for every cluster, so
\(M_l(G_l^\perp) = \operatorname{span}\{\mathbf{v}_l\}\) for every
\(l\).

\emph{Step 3 (own-cluster confinement and contradiction).} The same
column's own-cluster block, \(g_j \mathbf{v}_k + M_k e_j\), must lie in
\(M_k(G_k^\perp) = \operatorname{span}\{\mathbf{v}_k\}\); hence
\(M_k e_j \in \operatorname{span}\{\mathbf{v}_k\}\) for every item \(j\)
of cluster \(k\), so the entire range of \(M_k\) lies in
\(\operatorname{span}\{\mathbf{v}_k\}\). A symmetric matrix whose range
lies in a one-dimensional span is a multiple of the corresponding outer
product, so \(M_k = a_k\, \mathbf{v}_k \mathbf{v}_k'\); diagonality of
\(\Delta_k = M_k - \mathbf{b}_s^k \mathbf{b}_s^{k\prime}\) and Lemma A3
then force cluster \(k\) to be a spike or proportional, both excluded.
Hence no diagonal \(\Delta\) achieves rank \(K\), and
\(\operatorname{rank}(C+\Delta) \ge K+1\) for all diagonal \(\Delta\).
\(\square\)

\textbf{Remark (the mixed boundary).} When some clusters are
proportional, Lemma A4 makes \(r_k = 0\) available to a rival, the rank
budget acquires slack, and the alignment above is no longer forced: at
an exact rival, the non-proportional cluster's block can send
\(G_k^\perp\) along a direction \emph{tilted off} the span of its
restricted general column, a freedom the theorem's rank counting removes
but the mixed case retains. Consider the special case
\(M(G_k^\perp)\subseteq\operatorname{span}\{\mathbf{v}\}\) (writing
\(M\) and \(\mathbf{v}\) for that cluster's block \(M_k\) and restricted
general column \(\mathbf{v}_k\)), with
\(M = a\,\mathbf{v}\mathbf{v}' + \mathbf{v}\mathbf{z}' + \mathbf{z}\mathbf{v}'\)
and \(\mathbf{z} \perp \mathbf{v}\). This restricts the image of
\(G_k^\perp\), not the full image of \(M\), and already exhibits the
decisive algebra: diagonality of the corresponding \(\Delta\) reads
\(a + d_i + d_j = \rho_i \rho_j\) for \(i \ne j\), with
\(d_i = z_i/v_i\) and \(\rho_i = s_i/v_i\), and comparing two disjoint
index pairs \((i,m)\) and \((j,h)\) forces
\((\rho_i - \rho_m)(\rho_j - \rho_h) = 0\). A \emph{resistant} pattern
(two disjoint unequal ratio pairs) violates this identity, a three-item
cluster leaves it unconstrained, and a constant-except-one pattern
satisfies it. Because of the tilt freedom, the identity is not by itself
a proof in either direction. Table~\ref{tbl-boundary} examines selected
configurations: numerical minimization reproduces the analytically
constructed exact rivals for the three-item and constant-except-one
patterns to entry precision \(10^{-6}\) or better, while tested
resistant patterns retain positive local minima (\(D_3 = .001\)). A
complete characterization of the mixed case, including P3/P4-type
configurations with one proportional cluster beside resistant ones,
remains open; positive minima are evidence rather than proofs.

\begin{longtable}[]{@{}
  >{\raggedright\arraybackslash}p{(\linewidth - 6\tabcolsep) * \real{0.3725}}
  >{\raggedright\arraybackslash}p{(\linewidth - 6\tabcolsep) * \real{0.3529}}
  >{\raggedleft\arraybackslash}p{(\linewidth - 6\tabcolsep) * \real{0.1373}}
  >{\raggedleft\arraybackslash}p{(\linewidth - 6\tabcolsep) * \real{0.1373}}@{}}
\caption{Numerical checks of Theorem 1 and the mixed boundary: the
population distance \(D_K\) and the maximum absolute entry discrepancy
between \(\Sigma\) and the best-fitting \(K\)-factor model with free
diagonal uniquenesses.}\label{tbl-boundary}\tabularnewline
\toprule\noalign{}
\begin{minipage}[b]{\linewidth}\raggedright
Cluster configuration
\end{minipage} & \begin{minipage}[b]{\linewidth}\raggedright
Ratio structure
\end{minipage} & \begin{minipage}[b]{\linewidth}\raggedleft
\(D_K\)
\end{minipage} & \begin{minipage}[b]{\linewidth}\raggedleft
Max entry
\end{minipage} \\
\midrule\noalign{}
\endfirsthead
\toprule\noalign{}
\begin{minipage}[b]{\linewidth}\raggedright
Cluster configuration
\end{minipage} & \begin{minipage}[b]{\linewidth}\raggedright
Ratio structure
\end{minipage} & \begin{minipage}[b]{\linewidth}\raggedleft
\(D_K\)
\end{minipage} & \begin{minipage}[b]{\linewidth}\raggedleft
Max entry
\end{minipage} \\
\midrule\noalign{}
\endhead
\bottomrule\noalign{}
\endlastfoot
\emph{Every cluster non-proportional (Theorem 1)} & & & \\
P1 & 2+2 alternating, every cluster & .031 & .0099 \\
P2 & 2+2 alternating, every cluster & .004 & .0029 \\
Five 3-item clusters & three distinct ratios & .006 & .0069 \\
Five 4-item clusters & constant except one (3+1) & .002 & .0013 \\
\emph{Mixed proportional/non-proportional configurations (outside
Theorem 1)} & & & \\
P3 & 2+2 in four clusters, one silent & .017 & .0096 \\
P4 & 2+2 in four clusters, one silent & .002 & .0031 \\
3-item cluster & three distinct ratios & \(<10^{-8}\) & \(<10^{-6}\) \\
4-item cluster & constant except one (3+1) & \(<10^{-8}\) &
\(<10^{-7}\) \\
4-item cluster & 2+2 & .001 & .0146 \\
4-item cluster & four distinct (resistant) & .001 & .0129 \\
\emph{Reducible or correctly \(K\)-dimensional} & & & \\
P5 & proportional everywhere & \(<10^{-10}\) & \(<10^{-10}\) \\
P6 & \(K\)-factor structure by construction & \(<10^{-10}\) &
\(<10^{-10}\) \\
\end{longtable}

{\footnotesize\emph{Note.} These selected constructions illustrate how
within-cluster ratio patterns affect reducibility; they do not establish
a universal ratio-only criterion for the mixed boundary. Distance
magnitudes depend on the loading values and uniquenesses and illustrate
rather than bound separation. Both columns come from one minimization,
population-level ML factor analysis with at least 40 random starts
(tolerance \(10^{-8}\)): \(D_K\) is the minimized ML discrepancy between
\(\Sigma\) and the best-fitting \(K\)-factor model, formalized in
Section \ref{sec-estimand}, and Max entry is the largest absolute entry
of \(\Sigma - (\Lambda\Lambda' + \tilde\Psi)\) at the minimizer.
Near-zero residuals numerically check the fit of the analytically
constructed exact rivals; positive local minima are numerical evidence
against a rival, not proofs. The P rows are this paper's populations
(Section \ref{sec-s3-design} and Appendix B); the pattern rows are
constructed. Upper and lower panels \(K = 5\); middle single-cluster
rows \(K = 3\), one non-proportional cluster beside two proportional
ones. Loading values are those of the preliminary study's design
(Appendix B); the distances for this paper's six populations are in
Table \ref{tbl-sim1pop}. The P4 row uses the corrected declared weak
pattern; its isolated computational provenance and non-impact on the
simulation results are documented in the replication archive README.}

\textbf{Remark (attainment of \(D_K\)).} The minimum defining \(D_K\) in
Section \ref{sec-estimand} is attained. The function \(x-\log x-1\) is
coercive at both zero and infinity, so every finite \(F_{\mathrm{ML}}\)
sublevel bounds the generalized eigenvalues of a candidate covariance
relative to fixed \(\Sigma\) and is therefore contained in a compact
positive-definite covariance neighborhood. Writing a candidate as
\(C+\tilde\Psi\), with \(C\) positive semidefinite of rank at most \(K\)
and \(\tilde\Psi\) nonnegative diagonal, bounds both components
separately through their nonnegative diagonal entries. Every minimizing
sequence thus has a convergent subsequence; the nonnegative-diagonal
cone and the positive semidefinite rank-at-most-\(K\) set are closed, so
its limit remains admissible and attains the infimum. Where Theorem 1
excludes an exact \(K\)-factor representation, the attained value is
therefore strictly positive rather than a nonattained zero infimum.

\section*{Appendix B: Preliminary simulation
study}\label{appendix-b-preliminary-simulation-study}
\addcontentsline{toc}{section}{Appendix B: Preliminary simulation study}

\subsection*{Preliminary Stage A: estimation and comparison at the
generating structure}\label{sec-study1}
\addcontentsline{toc}{subsection}{Preliminary Stage A: estimation and
comparison at the generating structure}

\subsubsection*{Design}\label{sec-s1-design}
\addcontentsline{toc}{subsubsection}{Design}

All populations share \(K = 5\) clusters and \(J = 20\) items (four per
cluster, satisfying the parameter-identifiability minimum of Fang et al.
(\citeproc{ref-fang2021identifiability}{2021})). Responses are
multivariate normal and scaled to unit variance. Population loadings are
\emph{fixed} rather than randomly drawn: the treatment variable is the
loading pattern's proportionality structure, random draws would place
each replication at an uncontrolled distance from reducibility, and
fixed patterns pin the tier exactly and are exactly replicable.

Table~\ref{tbl-sim1pop} lists the six populations: five
bifactor/higher-order populations crossing the tier taxonomy with
group-loading magnitude, and one correlated-factors population serving
as the negative control.

\begin{longtable}[]{@{}
  >{\raggedright\arraybackslash}p{(\linewidth - 8\tabcolsep) * \real{0.0395}}
  >{\raggedright\arraybackslash}p{(\linewidth - 8\tabcolsep) * \real{0.1579}}
  >{\centering\arraybackslash}p{(\linewidth - 8\tabcolsep) * \real{0.0658}}
  >{\raggedright\arraybackslash}p{(\linewidth - 8\tabcolsep) * \real{0.6842}}
  >{\raggedleft\arraybackslash}p{(\linewidth - 8\tabcolsep) * \real{0.0526}}@{}}
\caption{Preliminary Stage A populations (all loadings fixed; \(K = 5\),
\(J = 20\)).}\label{tbl-sim1pop}\tabularnewline
\toprule\noalign{}
\begin{minipage}[b]{\linewidth}\raggedright
\#
\end{minipage} & \begin{minipage}[b]{\linewidth}\raggedright
Type
\end{minipage} & \begin{minipage}[b]{\linewidth}\centering
Gen.
\end{minipage} & \begin{minipage}[b]{\linewidth}\raggedright
Group loadings, by cluster
\end{minipage} & \begin{minipage}[b]{\linewidth}\raggedleft
\(D_5\)
\end{minipage} \\
\midrule\noalign{}
\endfirsthead
\toprule\noalign{}
\begin{minipage}[b]{\linewidth}\raggedright
\#
\end{minipage} & \begin{minipage}[b]{\linewidth}\raggedright
Type
\end{minipage} & \begin{minipage}[b]{\linewidth}\centering
Gen.
\end{minipage} & \begin{minipage}[b]{\linewidth}\raggedright
Group loadings, by cluster
\end{minipage} & \begin{minipage}[b]{\linewidth}\raggedleft
\(D_5\)
\end{minipage} \\
\midrule\noalign{}
\endhead
\bottomrule\noalign{}
\endlastfoot
P1 & Tier I, moderate & .70 & \((.3, .6, .3, .6)\) in every cluster &
.031 \\
P2 & Tier I, weak & .70 & \((.2, .4, .2, .4)\) in every cluster &
.004 \\
P3 & Tier II, moderate & .70 & cluster 1: \((.45, .45, .45, .45)\),
silent; clusters 2--5: \((.3, .6, .3, .6)\) & .017 \\
P4 & Tier II, weak & .70 & cluster 1: \((.30, .30, .30, .30)\), silent;
clusters 2--5: \((.2, .4, .2, .4)\) & .002 \\
P5 & Tier III (HO) & SL & first-order \(\lambda_j = .70\); second-order
\(\boldsymbol\gamma = (.75, .70, .65, .60, .55)\) & 0 \\
P6 & Oblique + cross & --- & primaries \(.70\); cross-loading \(.30\) on
the third item of each cluster (adjacent factor); \(\Phi = .30\) & 0 \\
\end{longtable}

{\footnotesize\emph{Note.} Alternating group values make a cluster
non-proportional; a constant group value against the constant general of
\(.70\) makes it exactly proportional (silent), with implied
second-order loading \(.841\) for the \(.45\) cluster and \(.919\) for
the \(.30\) cluster. P5 is exactly reducible by Proposition 1; P6 has no
general factor, with cross-loadings on non-anchor items. Communalities
stay below \(.90\). \(D_5\) = the population distance of Section
\ref{sec-estimand} to the best five-factor model, computed by
population-level ML factor analysis as in Table~\ref{tbl-boundary}; the
zeros are exact, by Proposition 1 for P5 and by construction for P6. The
P4 entry is the corrected boundary value whose isolated provenance and
scope are documented in the note to Table~\ref{tbl-boundary}.}

The populations' status under the theory is stated cluster by cluster.
P1 and P2 have every cluster non-proportional, so Theorem 1 applies
directly and both are distinguishable at the observed-covariance level.
P3 and P4 mix four resistant clusters with one proportional cluster and
therefore fall outside the theorem's hypothesis; their positive fitted
minima provide numerical evidence of irreducibility
(Table~\ref{tbl-boundary}), not a formal result. P5 is reducible by
Proposition 1, and P6 is a \(K\)-factor structure by construction. The
design crosses the six populations with \(N \in \{500, 1000, 3000\}\) at
200 replications per cell.

\subsubsection*{Fits and outcomes}\label{sec-s1-fits}
\addcontentsline{toc}{subsubsection}{Fits and outcomes}

Every replication receives the three anchored bifactor designs of
Section \ref{sec-notation} (anchor-only, anchor-zero, full-primary), all
built from the analyst's assumed five-cluster simple structure (general
column fully specified internally; \(K^* = 6\)), plus the matched
oblique-\(K\) model at each of the AZ and AO specifications, which
supplies the process's step-2 comparison. Applied to P6, the designs
encode the assumed clusters, not the truth; the cross-loadings sit on
non-anchor items, so AO and the oblique arm can recover them through
\(-1\) entries, AZ's anchor-row zeros are unaffected, and any joint
failure of all three on P6 is attributable to the misapplied bifactor
frame rather than to clashes between fixed zeros and true loadings.

\textbf{Step-2 comparison.} The anchored pair of Section \ref{sec-route}
is fitted at both specifications on the same replications with both
criteria recorded, so the choice of criterion and specification is
checked rather than assumed.

\subsubsection*{Results and findings}\label{sec-s1-results}
\addcontentsline{toc}{subsubsection}{Results and findings}

Table S2 in the online supplement carries the full estimation grid, and
the findings quote its decisive cells.

\textbf{Preliminary Finding 1: AO is slightly worse than but close to
AZ, and both are worse than full-primary.} The ordering appears only in
the hard cells (Table S2). At Tier-I moderate the three designs are
indistinguishable (minimum group congruence \(\ge .993\) at every
\(N\)). Under weak loadings at \(N = 500\), congruence runs \(.832\)
(AO), \(.907\) (AZ), and \(.968\) (full-primary); on the reducible P5,
\(.830\), \(.962\), and \(.991\). These gaps narrow markedly by
\(N = 3000\). Errors are omissions almost exclusively. The
false-discovery rate never exceeds \(.004\). Regularized selection can
therefore omit weak true loadings but almost never adds false ones. In
exploratory settings, AO is enough at moderate loadings; under weak
loadings at small \(N\), where its congruence falls near \(.83\) and its
false-negative rate approaches \(.38\), the anchored zeros of AZ buy
real protection.

Estimation quality and the general-factor decision also dissociate.
Reducible and wrong-frame populations estimate cleanly under AZ and
full-primary: P5 recovery at \(.96\) to \(.998\), though AO lags at
\(.830\) to \(.984\); P6 group columns at \(.97\) to \(.998\) beside a
spurious general. Yet the step-2 comparison concludes on both, in 100\%
of replications, that \(K\) dimensions suffice. Only the comparison
answers the general-factor question.

\textbf{Preliminary Finding 2: at the generating count, the operational
comparison has favorable operating characteristics in both directions.}
On the true-bifactor populations BIC favors the added dimension in a
large majority of replications at both specifications, the exceptions
confined to the weak populations at the smallest \(N\); on the
higher-order and oblique populations it favors the \(K\)-dimensional
representation in every replication (Table~\ref{tbl-s2step2depth}). On
the reducible population that is a parsimony decision among
covariance-equivalent representations rather than a rejection of the
higher-order reading (Section \ref{sec-disting}). Detection is
essentially complete by the largest \(N\) in every cell. The ELBO
columns track the same behavior and are somewhat more accurate in the
weakest small-\(N\) cell, but are retained as sensitivity results
because cross-design comparability is not established. These are
tentative operating characteristics for the selected anchored
comparison, not proof that it is an exact test of unrestricted \(D_K\).

\begin{longtable}[]{@{}
  >{\raggedright\arraybackslash}p{(\linewidth - 18\tabcolsep) * \real{0.0769}}
  >{\raggedleft\arraybackslash}p{(\linewidth - 18\tabcolsep) * \real{0.1026}}
  >{\raggedleft\arraybackslash}p{(\linewidth - 18\tabcolsep) * \real{0.1026}}
  >{\raggedleft\arraybackslash}p{(\linewidth - 18\tabcolsep) * \real{0.1026}}
  >{\raggedleft\arraybackslash}p{(\linewidth - 18\tabcolsep) * \real{0.1026}}
  >{\raggedleft\arraybackslash}p{(\linewidth - 18\tabcolsep) * \real{0.1026}}
  >{\raggedleft\arraybackslash}p{(\linewidth - 18\tabcolsep) * \real{0.1026}}
  >{\raggedleft\arraybackslash}p{(\linewidth - 18\tabcolsep) * \real{0.1026}}
  >{\raggedleft\arraybackslash}p{(\linewidth - 18\tabcolsep) * \real{0.1026}}
  >{\raggedleft\arraybackslash}p{(\linewidth - 18\tabcolsep) * \real{0.1026}}@{}}
\caption{Step-2 comparison at the true \(K\): mean margins and selection
percentages under both criteria (ELBO, BIC) at both specifications (AZ,
AO).}\label{tbl-s2step2depth}\tabularnewline
\toprule\noalign{}
\begin{minipage}[b]{\linewidth}\raggedright
Pop.
\end{minipage} & \begin{minipage}[b]{\linewidth}\raggedleft
\(N\)
\end{minipage} & \begin{minipage}[b]{\linewidth}\raggedleft
\(\Delta\)ELBO AZ
\end{minipage} & \begin{minipage}[b]{\linewidth}\raggedleft
\% bif
\end{minipage} & \begin{minipage}[b]{\linewidth}\raggedleft
\(\Delta\)ELBO AO
\end{minipage} & \begin{minipage}[b]{\linewidth}\raggedleft
\% bif
\end{minipage} & \begin{minipage}[b]{\linewidth}\raggedleft
\(\Delta\)BIC AZ
\end{minipage} & \begin{minipage}[b]{\linewidth}\raggedleft
\% bif
\end{minipage} & \begin{minipage}[b]{\linewidth}\raggedleft
\(\Delta\)BIC AO
\end{minipage} & \begin{minipage}[b]{\linewidth}\raggedleft
\% bif
\end{minipage} \\
\midrule\noalign{}
\endfirsthead
\toprule\noalign{}
\begin{minipage}[b]{\linewidth}\raggedright
Pop.
\end{minipage} & \begin{minipage}[b]{\linewidth}\raggedleft
\(N\)
\end{minipage} & \begin{minipage}[b]{\linewidth}\raggedleft
\(\Delta\)ELBO AZ
\end{minipage} & \begin{minipage}[b]{\linewidth}\raggedleft
\% bif
\end{minipage} & \begin{minipage}[b]{\linewidth}\raggedleft
\(\Delta\)ELBO AO
\end{minipage} & \begin{minipage}[b]{\linewidth}\raggedleft
\% bif
\end{minipage} & \begin{minipage}[b]{\linewidth}\raggedleft
\(\Delta\)BIC AZ
\end{minipage} & \begin{minipage}[b]{\linewidth}\raggedleft
\% bif
\end{minipage} & \begin{minipage}[b]{\linewidth}\raggedleft
\(\Delta\)BIC AO
\end{minipage} & \begin{minipage}[b]{\linewidth}\raggedleft
\% bif
\end{minipage} \\
\midrule\noalign{}
\endhead
\bottomrule\noalign{}
\endlastfoot
P1 & 500 & 86.5 & 100 & 43.4 & 100 & 343 & 100 & 442 & 100 \\
P1 & 1000 & 170.8 & 100 & 63.3 & 100 & 823 & 100 & 1102 & 100 \\
P1 & 3000 & 494.5 & 100 & 105.1 & 100 & 3090 & 100 & 4171 & 100 \\
P2 & 500 & 11.5 & 90.0 & 9.6 & 86.5 & 20 & 80.0 & 33 & 83.5 \\
P2 & 1000 & 33.6 & 100 & 22.8 & 98.0 & 175 & 100 & 296 & 100 \\
P2 & 3000 & 109.5 & 100 & 47.0 & 100 & 1197 & 100 & 2092 & 100 \\
P3 & 500 & 64.7 & 100 & 30.3 & 100 & 270 & 100 & 352 & 100 \\
P3 & 1000 & 132.9 & 100 & 45.0 & 100 & 660 & 100 & 895 & 100 \\
P3 & 3000 & 388.4 & 100 & 70.5 & 100 & 2403 & 100 & 3174 & 100 \\
P4 & 500 & 4.9 & 70.0 & 4.1 & 65.5 & 0.3 & 47.0 & 10 & 59.5 \\
P4 & 1000 & 22.2 & 99.5 & 13.6 & 92.0 & 128 & 100 & 216 & 100 \\
P4 & 3000 & 83.7 & 100 & 32.1 & 99.0 & 954 & 100 & 1671 & 100 \\
P5 & 500 & \(-14.2\) & 0 & \(-15.4\) & 0 & \(-52\) & 0 & \(-62\) & 0 \\
P5 & 1000 & \(-15.6\) & 0 & \(-18.5\) & 0 & \(-55\) & 0 & \(-76\) & 0 \\
P5 & 3000 & \(-19.9\) & 0 & \(-20.0\) & 0 & \(-63\) & 0 & \(-79\) & 0 \\
P6 & 500 & \(-14.8\) & 0 & \(-15.4\) & 0 & \(-60\) & 0 & \(-76\) & 0 \\
P6 & 1000 & \(-16.8\) & 0 & \(-17.9\) & 0 & \(-62\) & 0 & \(-91\) & 0 \\
P6 & 3000 & \(-20.5\) & 0 & \(-19.8\) & 0 & \(-76\) & 0 & \(-102\) &
0 \\
\end{longtable}

{\footnotesize\emph{Note.}
\(\Delta\mathrm{ELBO}=\mathrm{ELBO}_{\mathrm{bifactor}}-\mathrm{ELBO}_{\mathrm{oblique}}\)
and
\(\Delta\mathrm{BIC}=\mathrm{BIC}_{\mathrm{oblique}}-\mathrm{BIC}_{\mathrm{bifactor}}\),
so positive values favor bifactor under both margins. ``\% bif'' is the
percentage of replications favoring the bifactor under the adjacent
criterion and specification. At \(N = 3000\) the BIC detection side
reaches 99--100\% everywhere while the margins on P5 and P6 remain in
favor of the \(K\)-dimensional representation.}

\subsection*{Preliminary Stage B: sweeps, specifications, and
miscounts}\label{sec-study2}
\addcontentsline{toc}{subsection}{Preliminary Stage B: sweeps,
specifications, and miscounts}

Preliminary Stage A fitted every model at the generating five-cluster
structure. Stage B asks the questions that precede and follow that
assumption. How do the factor-structure sweeps behave? Do two design
choices the analyst must make, the anchoring of the backbone \(Q_0\) and
the anchoring at which the Step-2 comparison is posed, change the
answers? And what happens to the general-factor decision when the count
handed to it is wrong?

\subsubsection*{Design}\label{sec-s2-design}
\addcontentsline{toc}{subsubsection}{Design}

The populations, sample sizes, and estimation machinery are Stage A's.
Two arms are added.

\textbf{Sweeps and backbone anchoring.} Both modes of Section
\ref{sec-route} run on every population at every \(N\), launched from a
three-factor backbone anchoring the first three clusters with two
anchors each (\(K_0 = 3\); window \(K \in [3, 8]\); single-spike
\(v_0 = .001\)); within a replication the two modes receive the
\emph{identical} backbone matrix, so mode differences are not confounded
with specification, and the four selection rules of Section
\ref{sec-route} are recorded. Because the backbone is itself a
specification that no study has varied, the sweeps are also crossed with
AZ-type and AO-type backbones at \(N \in \{500, 1000\}\) on identical
data, a within-replication contrast whose two versions differ only in
the twelve fixed zeros the AZ specification places in anchor rows.

\subsubsection*{Results and findings}\label{sec-s2-results}
\addcontentsline{toc}{subsubsection}{Results and findings}

\begin{figure}

\centering{

\pandocbounded{\includegraphics[keepaspectratio]{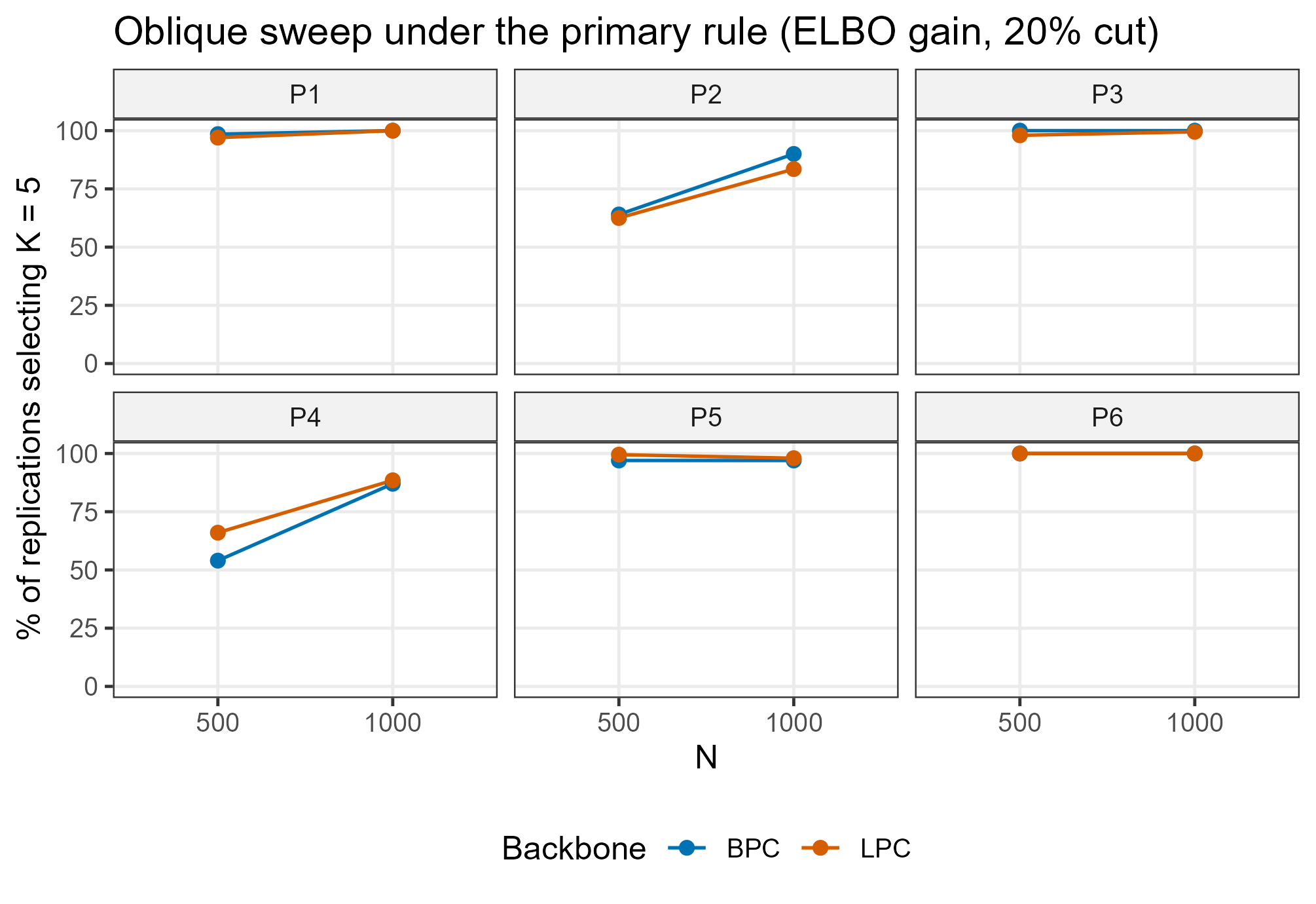}}

}

\caption{\label{fig-sweep20}Oblique sweep accuracy under the ELBO-gain
reading at the 20\% cut, by population, sample size, and backbone; 200
replications per cell. \(N = 3000\) was run only under the
over-restricted backbone of Preliminary Finding 4.}

\end{figure}

\textbf{Preliminary Finding 3: the oblique sweep counts better than the
bifactor sweep, under both backbones.} Figure~\ref{fig-sweep20} displays
the 20\% ELBO-gain reading by population, sample size, and backbone; the
executed grids are Tables S3 and S4 in the online supplement, carrying
the 20\% ELBO/BIC gain paths in oblique mode and raw criteria in both
modes and backbones.

In \emph{oblique} mode the 20\% ELBO-gain reading is the best overall
performer. Under AZ (Table S3) it is 98.5--100\% correct on P1/P3, 97\%
on P5, 100\% on P6, and 54--90\% on P2/P4; under AO (Table S4) the
corresponding ranges are 97--100\%, 98--99.5\%, 100\%, and 62.5--88.5\%.
The BIC-gain reading is less uniform (68--100\% under AZ; 18.5--100\%
under AO). Raw BIC is strongest on P2/P4 at \(N=500\) (77--88.5\%) but
nearly always over-counts P1/P3 (0--12.5\% correct across both sample
sizes), while raw ELBO lies between. The over-count has a principled
reading: on these irreducible populations the general dimension leaves a
positive gap at the core's count (Section \ref{sec-disting}), so a
rank-consistent criterion legitimately chases the additional common
dimension as information grows, while gain readings screen for the group
core by design. Both 20\% paths remain descriptive in the major study,
where truth-calibrated persistence determines delivery.

\textbf{Preliminary Finding 4: anchor-only (AO) and anchor-zero (AZ)
backbone sweeps are close.} The clean AZ and AO count paths show no
systematic winner: at P2, \(N=1000\), 20\% ELBO-gain accuracy is 90.0\%
versus 83.5\%, whereas P6 is 100\% under both backbones. A separate
over-restricted backbone, which added twelve correct zeros, drives the
gain rule to near-perfection on the moderate and reducible populations
while collapsing it on the weak populations. Thus selection depends
jointly on the rule, design, and population. The major developmental
study evaluates a different estimand--structural persistence--and finds
a clear AO advantage there; the online supplement records the
preliminary arm's audit history.

\textbf{Preliminary Finding 5: the oblique sweep can over-count by one,
and the stable structure remains in the studied populations.} On the
moderate and reducible populations, errors are upward only: a \(K + 1\)
solution retains the \(K\) structure beside a thin surplus column
(\citeproc{ref-chenjin2026fit}{Chen \& Jin, 2026a}). Table S5 uses the
preliminary 10\% gain screen to obtain enough \(K + 1\) solutions
because the 20\% rule leaves too few; the mechanism concerns those
solutions, not the screen. Matched columns remain highly congruent, with
blurring confined to weak-population item partitions. PID-5 shows a
similar adjacent-count pattern, but unknown truth makes it a boundary to
inspect rather than evidence of the mechanism.

\textbf{Preliminary Finding 6: the bifactor sweep can under-count, and
the stable structure fails with it.} Under bifactor mode, neither raw
criterion reaches 70\% correct in a clean AO/AZ cell. The 20\% ELBO-gain
rule reads 97--100\% in oblique mode but 0--26.5\% under AZ and 0--9\%
under AO. Errors run downward: P6 selects the absorbed \(K=4\)
essentially always, while P1's mean minimum truth congruence falls from
\(.98\) at \(K=5\) to \(.35\) at \(K=4\). Holzinger's lower-bound
bifactor count is a descriptive analogy only.

The opposite error is costly. At \(K=4\), merging clusters 4 and 5 makes
bifactor win every P5 and P6 replication although neither has a
separable general dimension; the contrast grows with \(N\) to \(+1730\)
and \(+2688\), respectively. The general column carries merged
covariance without fit warning, so general-factor inference remains
count-conditional.

\end{document}